\documentclass[a4paper,onecolumn,11pt,accepted=2022-05-31]{quantumarticle}
\pdfoutput=1
\usepackage[utf8]{inputenc}
\usepackage[english]{babel}
\usepackage[T1]{fontenc}
\usepackage{amsmath}
\usepackage{hyperref}

\usepackage{tikz}
\usepackage{lipsum}

\usepackage{indentfirst}
\usepackage{amsfonts}
\usepackage{amsmath}
\usepackage{amssymb}
\usepackage{mathtools}
\usepackage{bm}
\usepackage{graphicx}

\usepackage{etoolbox}
\usepackage{color}
\usepackage{booktabs}

\usepackage{dsfont}
\usepackage[numbers,sort&compress]{natbib}
\usepackage{dcolumn}   
\usepackage{bm}        
\usepackage[mathscr]{eucal}

\makeatletter

\usepackage{color}

\usepackage{bm}        

\definecolor{myblue}{rgb}{0,0,0.75}

\usepackage{soul}
\usepackage{ulem} 

\newcommand{\ep}{\varepsilon}

\newcommand{\be}{\begin{equation}}
\newcommand{\ee}{\end{equation}}
\newcommand{\bem}{\begin{multline}}
\newcommand{\eem}{\end{multline}}
\def\ba{\begin{aligned}}
\def\ea{\end{aligned}}
\newcommand{\bea}{\begin{eqnarray}}
\newcommand{\eea}{\end{eqnarray}}
\def\bes{\begin{subequations}}
\def\ees{\end{subequations}}
\def\bal{\begin{align}}
\def\eal{\end{align}}

\newcommand\bra[1]{\ensuremath{\langle#1|}}
\newcommand\ket[1]{\ensuremath{|#1\rangle}}
\newcommand\braket[2]{\ensuremath{\langle#1|#2\rangle}}
\newcommand\bracket[3]{\ensuremath{\langle#1|#2|#3\rangle}}
\newcommand\mean[1]{\ensuremath{\langle#1\rangle}}
\newcommand\abs[1]{\ensuremath{\left|#1\right|}}

\newcommand\lrp[1]{\left(#1\right)}
\newcommand\lrb[1]{\left[#1\right]}
\newcommand\lrv[1]{\left|#1\right|}
\newcommand\subs[1]{\left.#1\right|}
\newcommand\caseS[1]{\left\{#1\right.}
\newcommand{\lra}{\quad \Leftrightarrow \quad}

\newcommand{\lv}{\left|}
\newcommand{\rv}{\right|}
\newcommand{\la}{\left\langle}
\newcommand{\ra}{\right\rangle}

\newcommand{\lb}{\left[}
\newcommand{\rb}{\right]}

\renewcommand{\Re}{{\rm \, Re\,}}
\renewcommand{\Im}{{\rm \, Im\,}}

\pdfoutput=1

\begin{document}

\title{Non-ergodic delocalized phase with Poisson level statistics}

\author{W.~Tang}
\affiliation{Department of Physics, University of California, Berkeley, California 94720, USA}

\author{I.~M.~Khaymovich}
\email{ivan.khaymovich@gmail.com}
\homepage{https://sites.google.com/view/ivan-khaymovich/}
\orcid{0000-0003-2160-5984}
\affiliation{Max-Planck-Institut f\"ur Physik komplexer Systeme, N\"othnitzer Stra{\ss}e~38, 01187-Dresden, Germany}
\affiliation{Institute for Physics of Microstructures, Russian Academy of Sciences, 603950 Nizhny Novgorod, GSP-105, Russia}
\affiliation{Nordita, Stockholm University and KTH Royal Institute of Technology Hannes Alfv\'ens v\"ag 12, SE-106 91 Stockholm, Sweden}
\maketitle

\begin{abstract}{
Motivated by the many-body localization (MBL) phase in generic interacting disordered quantum systems,
we develop a model simulating the same eigenstate structure like in MBL, but in the random-matrix setting.
Demonstrating the absence of energy level repulsion (Poisson statistics), this model carries non-ergodic eigenstates,
delocalized over the extensive number of configurations in the Hilbert space.
On the above example, we formulate general conditions to a single-particle and random-matrix models in order to carry such states, based on the transparent
generalization of the Anderson localization of single-particle states and multiple resonances.
}
\end{abstract}

%

\vspace{10pt}
\noindent\rule{\textwidth}{1pt}
\tableofcontents\thispagestyle{fancy}
\noindent\rule{\textwidth}{1pt}
\vspace{10pt}



\section{Introduction}

The phenomenon of quantum thermalization in generic isolated quantum systems is highly non-trivial.
Indeed, the unitarity of quantum evolution keeps all the information about the initial state (unlike the dynamics in classical chaotic systems).
Therefore the relaxation of isolated quantum systems to the equilibrium thermal state, independently of the initial configuration,
gives rise to an obvious paradox.
This paradox is partially resolved by the ergodicity hypothesis~\cite{Deutsch1991,Srednicki1994,Srednicki1996,rigol2008thermalization,Polkovnikov_2011,DAlessio2016ETH}, claiming that as time goes, the quantum information diffusively scrambles among extensive number of highly non-local degrees of freedom and, thus, most of physical (local) observables acting on a part of the system thermalize, because the rest of the system plays a role of a bath for them.

However, as now accepted in the literature, quantum thermalization may generically fail as one applies a strong enough disorder to the system.
This happens due to the emergence of local integrals of motion and leaves a way for a new phase of matter, called many-body localization (MBL)~\cite{Basko06,gornyi2005interacting,Pal2010,oganesyan2007localization,Alet_CRP}.
The presence of an extensive number of the local integrals of motion~\cite{serbyn2013local,huse2014phenomenology} in the MBL phase
prevents the quantum information scrambling and also localizes the (particle) transport across the system.
In addition, the MBL phase is characterized by the Poisson level statistics (absence of level repulsion),
while many-body wave functions, localized in the real space, occupy a zero fraction, but extensive number of Hilbert-space configurations~\cite{Luitz15,Tikhonov2018MBL_long-range,Mace_Laflorencie2019_XXZ,luitz2019multifractality,QIsing_2021}~\footnote{There are many other physical quantities
characterizing the ergodicity breaking like fluctuations of local observables~\cite{Deutsch1991,Srednicki1994,Srednicki1996}, entanglement entropy~\cite{Bardarson_Sent_log(t),Abanin_RMP,Sent2020_Haque}, out-of-time-ordered correlators~\cite{Fan_OTOC_MBL,Huang_OTOC_MBL} and operator spreading~\cite{Abanin_RMP}, showing deviations from their ergodic values (sometimes differently from other observables~\cite{Fradkin_Sq_Dq,DeTomasi_Sq_Dq_2020}). In this paper, instead, we focus on those of these quantities which have a clear meaning in the random-matrix setting, namely, eigenstate localization properties in the Hilbert space and eigenlevel statistics.}.
Such states bear the name of non-ergodic extended or {\it multifractal} wave functions in the Hilbert space.

The complexity of the above mentioned many-body problems prevents one from analytical description of them. Thus, it is of particular concern and high demand to model essential attributes of these phenomena in a universal manner. In this respect, the random-matrix setting provides a perfect and more tractable playground, which can be used as quantum simulations of many-body systems.
In the quantum thermalizing phase, the standard Gaussian random matrices~\cite{Mehta2004random} with the Wigner-Dyson level repulsion do their job.
However going to the non-ergodic (especially MBL) phase, we face an immediate problem as the overwhelming majority of the random-matrix ensembles shows multifractal properties in the Hilbert space
only at the very point of the Anderson transition~\cite{Evers2008anderson}. There are only few ensembles like the Rosenzweig-Porter model~\cite{RP,Kravtsov_NJP2015} or some related ones~\cite{Nosov2019correlation,Nosov2019mixtures,LN-RP_RRG,LN-RP_WE,LN-RP_dyn,BirTar_Levy-RP,Kutlin2021emergent,kutlin2023anatomy,Motamarri2021RDM,DeTomasi2022nonHermitian}
which provide an entire phase of non-ergodic delocalized eigenstates.
However, all such models show the standard Wigner-Dyson level repulsion in the whole non-ergodic phase, like in Gaussian random matrix ensembles~\footnote{There are also models based on the quasiperiodic potential instead of uncorrelated on-site disorder which do show the deviations from the Wigner-Dyson level repulsion of the non-ergodic extended states~\cite{Floquet_MF,Sarkar_Floquet_MF,Ray_Floquet_MF,Sarkar2021signatures}, but unfortunately they also do not show Poisson level statistics either.}.

The origin of the non-ergodic Hilbert-space delocalization of MBL wave functions is at least two-fold.
First, the structure of the Hilbert space matters. The number of available configurations $N$ in a generic quantum system scales exponentially with the system size $L$, i.e. $N\sim e^{c L}$, and even if each single particle is localized in the coordinate space, it can still spread over few lattice sites there corresponding to a finite localization length $\xi$. It is this finiteness of $\xi>0$, which leads to the delocalization in the Hilbert space.
Indeed, one can estimate the number of configurations occupied by the $M$-particle wave-function with a finite filling fraction $\nu = M/L$ as $\xi^M \sim N^{D}$, with $D\simeq(\nu/c)\ln \xi>0$ providing the above non-ergodic delocalization (see, e.g.,~\cite{Mace_Laflorencie2019_XXZ,QIsing_2021}).
Second, the correlated nature of the diagonal matrix elements in the Hilbert-space configurations (given by the random uncorrelated on-site disorder and few-particle diagonal interactions in the coordinate space) leads to a rather complicated structure of multiple resonances~\cite{Burin_AdP,BurinPRB2015-1} (unlike the case of the Anderson localization), which, in turn, delocalizes the many-body wave function in the Hilbert space.

To sum up, due to the presence of the Hilbert space in addition to the coordinate space, MBL provides an example of wave functions with Poisson level statistics, which are (non-ergodically) extended in the Hilbert space.
Such a combination is typically absent in the picture of the Anderson localization or in any random-matrix ensemble.

In this paper we provide such an example in random-matrix setting and unveil its main ingredients, summarized below~\footnote{Here and further we associate the Hilbert space of a many-body system and a random-matrix ensemble by mapping their matrix indices and making the Hilbert space dimension $N$ equal to the matrix size.}.
\begin{enumerate}
  \item \label{item:j_N^gamma/2} In order to realize the structure of multiple resonances similar to MBL, one has to make the ratio of the amplitude of the hopping term $j$ to the disorder amplitude $W$ scaling up $j/W \sim N^{\gamma/2}$, $0<\gamma<1$, with the matrix size $N$.~\footnote{In some models, including the one which we consider, one can instead just consider the eigenenergy to be scaling with the size, $E\sim N^{\gamma/2}$.}
  \item \label{item:rare_resonances} The Poisson level statistics can be realized if resonant hopping terms between sites $i$ and $j$, satisfying the inequality $|j_{ij}|>|\ep_i-\ep_j|$,  form either an effectively block diagonal matrix with the extensive number of blocks $N_B\gg 1$, or are only few per row and only at the large, but sub-extensive distance $1\ll R\ll N$ from the site $i$.
\end{enumerate}

An immediate straightforward example satisfying the above conditions can be realized in a one-dimensional Anderson model by the scaling up ratio $j/W \sim N^{\gamma/2}$.
Indeed, as soon as the wave functions exponentially decay on the scale of the localization length $\xi \sim (j/W)^2 \sim N^{\gamma}$, which, in turn, scales sub-extensively with the system size, $0<\gamma<1$, the eigenstates are non-ergodic and delocalized,
while as soon as all such states are effectively split into a sub-extensive number $N_B\sim N/\xi = N^{1-\gamma}$ of blocks of independent eigenvalues
the corresponding level statistics should be Poisson.
This simple example provides some intuition of how the MBL eigenstate structure can be mimicked, however, in somehow trivial manner.
In the following, we will focus on a more realistic models, relevant to many-body systems and experimental realizations.

In terms of the many-body systems, the above ingredients~\ref{item:j_N^gamma/2},~\ref{item:rare_resonances} find applications in physically relevant random-matrix models~\cite{LN-RP_RRG,LN-RP_WE,LN-RP_dyn}.
Indeed, as soon as one considers the Hilbert-space structure of a generic many-body system~\cite{Tarzia_2020,QIsing_2021,Huse21},
one immediately sees that the main contribution to the many-body delocalization is given by the Hilbert-space configurations separated by the extensive (Hamming) distance $\sim L\sim\ln N$.
The statistics of the corresponding matrix elements, coupling such configurations, appears to have two main properties related to the above mentioned ingredients:
First, the typical matrix elements should scale as a power of the Hilbert-space dimension $N$ (like in the item~\ref{item:j_N^gamma/2}), and,
second, the distribution of them should be fat-tailed (at least log-normal~\cite{LN-RP_RRG,LN-RP_dyn,Huse21}), leading to rare resonances (like in the item~\ref{item:rare_resonances}).

From the experimental point of view, the random-matrix models  provide an excellent description of the Anderson transition in $d$-dimensional dipolar systems.
Indeed, since $1989$ it is known that in the random-matrix setting one can realize an Anderson transition at any $d$-dimensional lattice by considering
a generalized dipolar hopping term (like in~\cite{Levitov1989,Levitov1990,MirFyod1996}) with the exponent $a$ of the power-law decay $j_{m,m+R}\sim 1/R^a$.
For large $a>d$ all the states are known to be (power-law) Anderson localized, while for small enough $a<d$ they become delocalized and ergodic, and
this transition is barely affected by any finite amplitude $W$ of on-site disorder.

However, the latter delocalization property at small powers $a<d$ of power-law long-range systems might fail in the dipolar systems,
where all the dipoles are not randomly oriented~\cite{Levitov1989,Levitov1990}, but aligned, e.g., by an electric field~\cite{Burin1989,Malyshev2005,Deng2018duality,Nosov2019correlation,Kutlin2020_PLE-RG,deng2020anisotropy}.
In the literature this is called the Burin-Maksimov model by the names of the authors of~\cite{Burin1989} first suggested it.
In this case, the localization scenario is different: for all $a<3d/2$ at moderate disorder $W$ (and for all $W$ at $a<d$) there are high-energy states
which are delocalized and ergodic, but for $d\leq 2$ their fraction in thermodynamic limit, $N\to\infty$, is zero among the entire spectrum~\cite{Malyshev2005}~\footnote{The presence of this measure zero of ergodic states mimics more the many-body mobility edge, present just before the MBL transition, than the MBL phase itself, however in the former case the mid-spectrum ergodic states form a measure one of all the eigenstates.
Similarity is more in the divergence of the energies of these ergodic states with respect to the MBL ones: like in the considered Burin-Maksimov model
these energies grow with the system size with respect to the ground state energy. Therefore for any initial state, with the finite temperature (or equivalently finite energy density) below the many-body mobility edge, the localization will survive in the thermodynamic limit.}.
Unlike that measure zero of ergodic states, the spectral-bulk eigenfunctions are all power-law localized for {\it any} $a>0$ with the effective power-law decay exponent being symmetric with respect to $a=d$~\cite{Deng2018duality,Nosov2019correlation}~\footnote{This localization, symmetric with respect to $a=d$, survives also under the positional disorder~\cite{Kutlin2020_PLE-RG} and finite anisotropy~\cite{deng2020anisotropy}, however it is fragile to the reduction of correlations~\cite{Nosov2019mixtures,Kutlin2021emergent} and mutual disorientation of different dipoles~\cite{Levitov1989,Levitov1990}}.

\begin{figure}[t]
  \center{
  \includegraphics[width=0.4\textwidth]{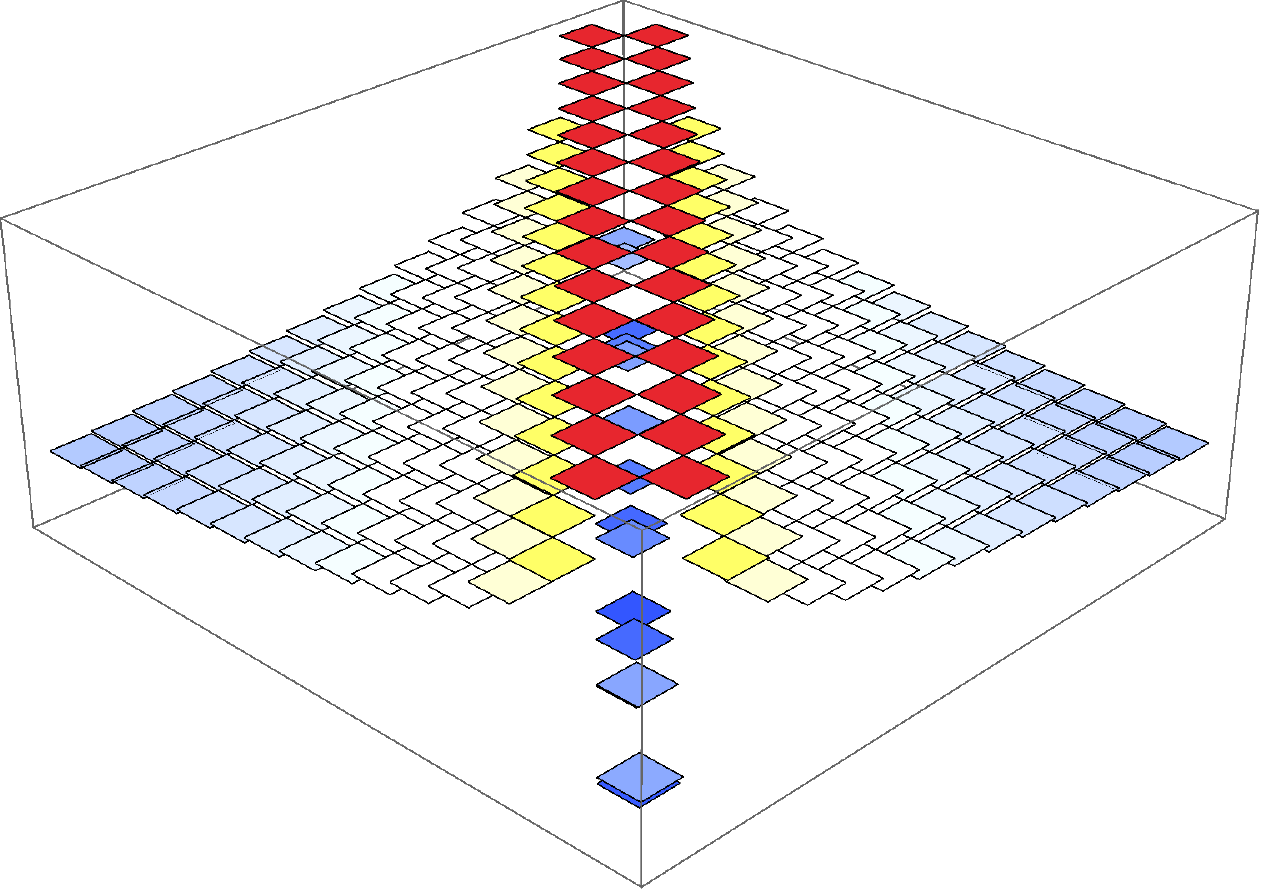}
  \includegraphics[width=0.33\textwidth]{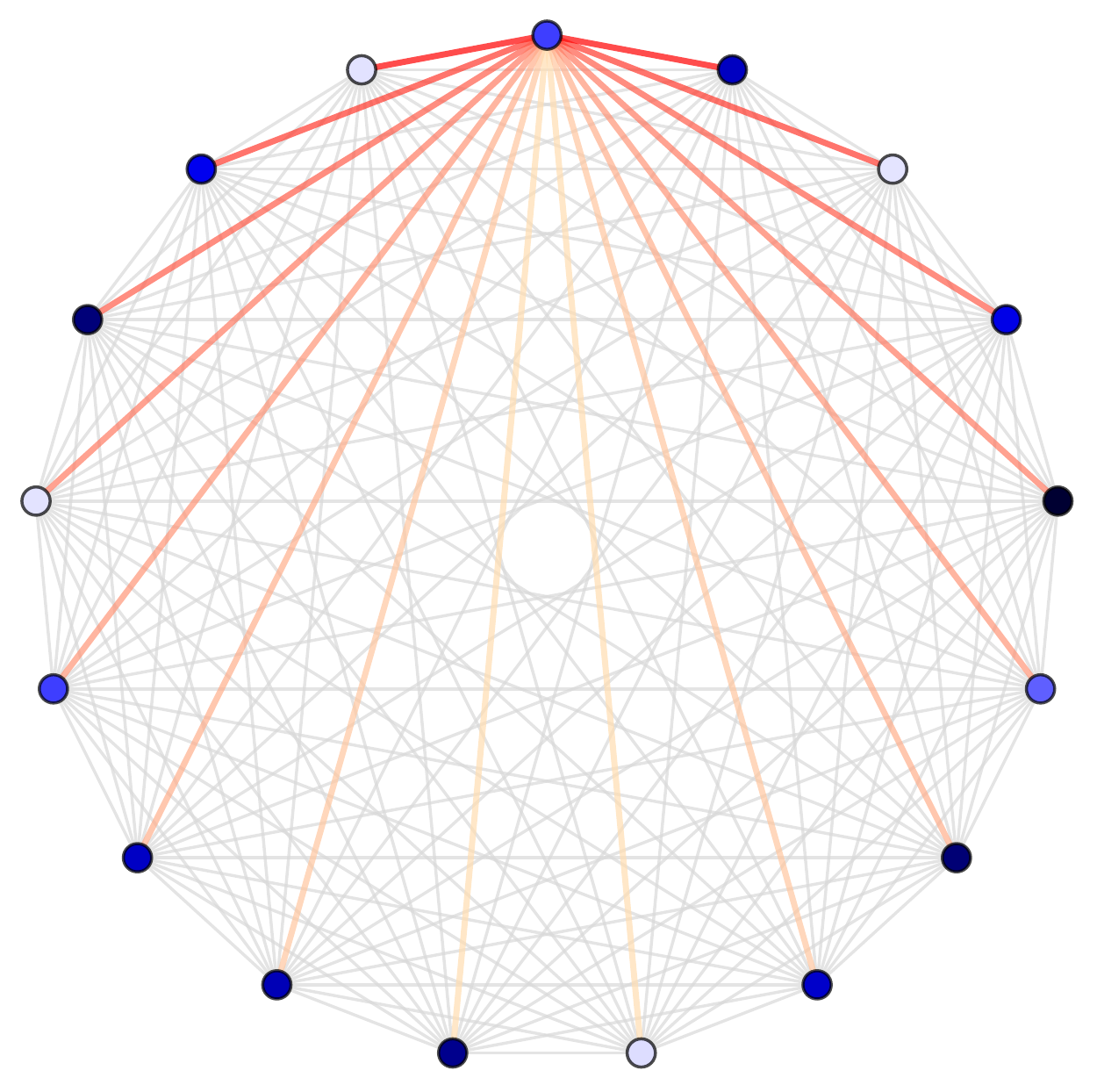}
  }
  \caption{\textbf{Sketch of the generalized Burin-Maksimov model} defined in Eqs.~\eqref{eq:ham_mat}-\eqref{eq:dist}.
  The brightness of blue-colored diagonal elements~(left) and vertices~(right) corresponds to the on-site disorder $\ep_n$,
  while the color from red to yellow and light blue of the off-diagonal elements~(left) and edges~(right) shows the power-law decay $\sim \frac{N^{\gamma/2}}{||m-n||^a}$ of hopping from/to the topmost vertex.
  In both panels we plot a realization of the matrix for $a=0.5$, $\gamma=1$, and $N=17$.
  \label{Fig:distance_sketch}
  }
\end{figure}

Already this very Burin-Maksimov (BM) model, Fig.~\ref{Fig:distance_sketch}, places the question of the presence of non-ergodic extended states with Poisson statistics.
Indeed, the localized bulk states, showing Poisson statistics, are separated from the spectral edge ergodic ones, which have spectral degeneracies and extensive energies scaling with the matrix size. This hints that the states in the buffer between the bulk ($E\sim O(1)$) and the edge ($E\sim N^{d-a}$) spectral states {\it may} combine the properties of both: they can be non-ergodic, delocalized, and demonstrate Poisson statistics at the same time.
Unfortunately, they form again just a zero fraction of all the states in the spectrum.
In order to overcome the last drawback of BM model, in this paper we consider its modification,
with the bulk majority of eigenstates being non-ergodic, delocalized, and showing Poisson statistics
by utilizing the first of the above ingredients, namely, by making the hopping amplitude scaling as $j/W \sim N^{\gamma/2}$ with respect to the on-site disorder.

Namely, we consider the above Burin-Maksimov model determined by the $N\times N$ matrix, Fig.~\ref{Fig:distance_sketch}(left)
\be\label{eq:ham_mat}
H_{mn} = \delta_{mn} \ep_n + j_{mn} \ , \text{ with } j_{mn} = \frac{t}{||m-n||^a} \ ,
\ee
integers $0\leq m,n < N$, and $N$ independently identically distributed random numbers $\ep_n$ with zero mean and the following variance
\be
\mean{\ep_n} = 0 \ , \quad \mean{\ep_n^2} = W^2 \ .
\ee
To satisfy the condition of multiple resonances we use the ingredient~\ref{item:j_N^gamma/2} and scale the hopping amplitude $t$ with the matrix size $N$
\be
t = N^{\gamma/2} \ .
\ee
For simplicity of analytical consideration we consider periodic boundary conditions and determine the distance $||m-n||$ as the minimal length on the one-dimensional (1d) circle, see Fig.~\ref{Fig:distance_sketch}(right):
\be\label{eq:dist}
||m-n|| = \min(|m-n|, |N+n-m|) \ .
\ee

\section{Known methods}\label{Sec:methods}
In this section, we remind several methods to determine localization known in the literature and apply it to the above model (focusing mostly on $\gamma=0)$.
We consider the methods written in the papers~\cite{Levitov1989,Levitov1990,Nosov2019correlation}.
Those who is interested in the results and current analytical consideration, can go directly to Secs.~\ref{Sec:Results} and~\ref{Sec:analytics}.

\subsection{Anderson's resonance counting}\label{sec:ALT}
Here we count the number of sites $m$ which are in resonance, $|\ep_n-\ep_m|<j_{mn}$, with a certain $n$th one.
As soon as this number of resonances is finite, the state $n$ is Anderson localized.
Similarly to~\cite{Levitov1989}, we count the resonances at each distance $||m-n||\sim R$ where in 1d the level spacing $\ep_n - \ep_m$ between closest levels is given by
$\delta_R \simeq W/R$, while $j_{m,m+R} = N^{\gamma/2}/R^a$: if $j_{m,m+R}>\delta_R$, the number of resonances will scale as $\sim R$.

As one can see from above expressions, the number of resonances is finite only if most of resonances appear at small distances $R\lesssim O(1)$, i.e., for $a>1$ and $\gamma\leq 0$,
or don't appear at all, $j_{m,m+R}<\delta_R$ for $1\leq R<N$, i.e., for $\gamma\leq-2$.

For the other parameters, the number of resonances formally diverges with increasing $N$.
However, unlike the random (uncorrelated) models~\cite{Levitov1989,Levitov1990,MirFyod1996}, the resonance counting fails to describe the wave-function structure in this case. The origin of this failure follows from the fact that this divergence in the models with the correlated kinetic terms leads to the delocalization of a measure zero of eigenstates (or even one state in Richardson model) that take the most weight of this kinetic term due to their high energies (for the details please see~\cite{Nosov2019correlation}).
Therefore in the next subsections we consider another method invented in~\cite{Nosov2019correlation} and called the matrix-inversion trick.

\subsection{Momentum-space localization and matrix-inversion trick}
The matrix~\eqref{eq:ham_mat} corresponds to the Hamiltonian
\be\label{eq:ham}
\hat H = \hat \ep+\hat j \
\ee
written in the coordinate basis $\{\lv n \ra\}$
via the diagonal disorder
\be
\hat \ep = \sum_n \ep_n \lv n \ra \la n \rv
\ee
and the translation-invariant hopping term
\be\label{eq:j_mn}
\hat j = \sum_{m,n} j_{mn}\ket{m}\bra{n} = \sum_{q} j_{q}\ket{q} \bra{q} \ .
\ee
The latter can be diagonalized in a momentum basis
\be
\ket{q} = \sum_{n=0}^{N-1} \frac{e^{i q n}}{\sqrt{N}}\ket{n} \ ,
\ee
as the translation invariance of the hopping term $j_{mn}=j_{||m-n||}$ and periodic boundary conditions, Fig.~\ref{Fig:distance_sketch}(right), allow one to find the spectrum of $\hat j$ via the Fourier transform
\be
j_q = \sum_n j_n e^{i q n} \ ,
\ee
with $q = 2\pi k/N$ and integer $0\leq k<N$. Here and further all the summations over $n$ and $m$ are taken over the entire interval $0\leq m,n<N$.
General results for the spectrum $j_q$ given, e.g., in~\cite{Nosov2019correlation} take the form
\be\label{eq:j_q}
\frac{j_q}{t} = \left\{
        \begin{array}{ll}
          A_a q^{a-1} + \zeta_a, &\quad |q|\ll 1, \\
          C_\pi + B_a|\pi-q|^{2}, &\quad |\pi - q|\ll 1,
        \end{array}
      \right.
\ee
Here $A_a = \Gamma_{1-a} \sin \frac{\pi a}{2}$, $B_a = 2(1-2^{3-a})\zeta_{a-2} \simeq a/2$, $C_\pi = 2(2^{1-a}-1)\zeta_a$,
$\zeta_a$ and $\Gamma_a$ are the zeta and Gamma functions.
Here and further we restrict our consideration to the range $0<a<2$ for simplicity where the maximal value of $j_q$ is given by
\be
\frac{j_{q=0}}{t} \sim  \zeta_a + \frac{N^{1-a}}{1-a}
\ee
and the minimum is negative and equal to $C_\pi$~\footnote{Further for simplicity we shift both the energy $E$ and the hopping term with respect to its minimal value $C_\pi$.}.

In the momentum basis where $j_q$ plays a role of the diagonal potential, the disorder $\hat \ep$ transforms to a translation-invariant hopping term
\be
\hat \ep = \sum_{p,q} \ep_{p-q} \ket{p}\bra{q} \ , \text{ with }
\ep_{q} = \sum_n \frac{e^{i q n}}{N} \ep_n, \text{ and }  \mean{|\ep_q|^2} = \frac{W^2}{N} \ .
\ee

Further in this part, we consider the localization of the states in the momentum space due to the large diagonal elements $j_q$ of the high-energy spectral edge states
followed by the method to understand why the presence of such states may localize the spectral bulk states in the coordinate basis.

\subsubsection{Momentum-space localization}
Similarly to the previous section~\ref{sec:ALT}, one can count the number of resonances in the momentum space, but taking into account the fact that
the level spacing $\delta j_q = |j_q - j_{q+\pi/N}|$ is highly $q$-dependent
\be\label{eq:j_q}
\frac{\delta j_q}{t} = \left\{
        \begin{array}{ll}
          |(a-2)A_a| q^{a-2}/N, &\quad |q|\ll 1, \\
          2 |B_a(\pi-q)|/N, &\quad |\pi - q|\ll 1,
        \end{array}
      \right. \ .
\ee
It is given by
\be
\delta j_{q\simeq 1} \sim \frac{t}{N}
\ee
in the bulk $q\simeq 1$, while the ones at the edges of the $q$-range take the form
\be
\delta j_{q = \pi-\pi/N} \sim \frac{t}{N^2}, \quad \delta j_{q = 0} \sim \frac{t}{N^{a-1}} \ .
\ee
Thus, the minimal level spacing is always (in our range of interest, $0<a<2$) given by $q=\pi$, while the maximal is $\delta j_{0}$. 

Taking all this into account, one can find at which $q$s the states are localized in the momentum space (therefore mostly represented by the plane waves in the coordinate basis):
\be
|\delta j_q|>|\ep_q| \sim N^{-1/2}
\Leftrightarrow
\caseS{
q^{a-2}>N^{(1-\gamma)/2} \ , \quad a<2 \ , \gamma<1\atop
\text{most } q \ , \quad \gamma>1 \ .
}
\ee
As we are not interested in the localization in the momentum space of most (measure one of all) bulk states, further we focus on $\gamma<1$ and $0<a<2$.
In this case only zero fraction $q^*$, of eigenstates are localized in the momentum basis, i.e. nearly plane waves in the coordinate basis
\be\label{eq:q^*}
|q|<q^* = N^{-(1-\gamma)/[2(2-a)]}\ll O(1) \ , \quad \gamma<1 \ , 0<a<2 \ .
\ee
Moreover they have an extensive energy for $a<1$
\be
E \simeq j_{|q|<q^*} > N^{(\gamma-1)(1-a)/[2(2-a)]} \gg O(1) \ .
\ee
These factors together are necessary for a so-called correlation-induced localization~\cite{Nosov2019correlation}, when a measure zero of high-energy states causes the localization of the bulk spectrum states beyond the convergence of the resonance counting.

\subsubsection{Correlation-induced localization and matrix inversion trick}\label{sec:MI}
The presence of (measure zero of) high-energy states, delocalized in the coordinate basis, leads to the reduction of the effective hopping term for the bulk spectral states~\cite{Borgonovi_2016,Nosov2019correlation,deng2020anisotropy,Motamarri2021RDM}.
Indeed, on one hand, using the localization of the high-energy states $|q|\ll q^*$ in the momentum space, one can find the exact eigenstates with the energy $E_q \simeq j_q$ of the entire Hamiltonian~\eqref{eq:ham}
perturbatively
\be\label{eq:psi_q_perturb}
\ket{\psi_q} = \ket{q} + \sum_{p\ne q} \frac{\ep_{q-p}}{j_q - j_p}\ket{p} + \ldots
\ee
Thus, the hopping term~\eqref{eq:j_mn} can be rewritten using the above eigenstates as
\be
\hat j = \sum_{|q|<q^*} j_{q}\ket{\psi_q} \bra{\psi_q} + \hat j_{\rm eff} \ ,
\ee
where due to the divergence of eigenenergies~\eqref{eq:j_q} at $a<1$ the first term gives the main contribution, while the term $j_{\rm eff}$,
corresponding to the summation over the states, ergodic in the momentum space, $|q|>q^*$, and to the perturbative terms from Eq.~\eqref{eq:psi_q_perturb},
is drastically reduced with respect to the initial one. In a sense, $j_{\rm eff}$ corresponds to the effective hopping term in the Hilbert space sector, orthogonal with respect to the high-energy states, which are non-ergodic in the momentum space.
On the other hand, as $\ket{\psi_q}$ are the exact eigenstates of the entire Hamiltonian~\eqref{eq:ham} they should be orthogonal, $\braket{\psi_q}{\psi_E}=0$, to all other eigenstates $\ket{\psi_E}$.
Therefore the effective hopping for $\ket{\psi_E}$ is just $j_{\rm eff}$
\be
\hat j\ket{\psi_E} = \sum_{|q|<q^*} j_{q}\ket{\psi_q} \braket{\psi_q}{\psi_E} + \hat j_{\rm eff}\ket{\psi_E} = \hat j_{\rm eff}\ket{\psi_E} \ .
\ee
This already hints that these bulk eigenstates can be localized in the coordinate space even beyond localization perturbation theory of Sec.~\ref{sec:ALT} due to the presence of correlations in the hopping term $j_{mn}$.

In order to estimate $j_{\rm eff}$, in~\cite{Nosov2019correlation} the authors invented a so-called matrix-inversion trick~\footnote{This method has recently been generalized in~\cite{Motamarri2021RDM} in order to describe non-ergodic delocalized states as well. Alternative methods have been considered in~\cite{Kutlin2021emergent}.}
Here we take the freedom to remind it to the reader.
The main idea behind the matrix-inversion trick is to get rid of the divergence in the spectrum $j_q$~\eqref{eq:j_q} by inverting the hopping matrix.
Before that, one should shift $\hat j$ by an identity multiplied by a certain scalar $E_0 \sim O(1)$ in order to not only make former large-energy terms small,
but also to avoid adding some large-energy terms to the inverted model due to the possible resonances:
\begin{multline}
E \ket{\psi_E} = \lrp{\sum_q j_q\ket{q}\bra{q}  + \sum_n \ep_n \ket{n} \bra{n} }\ket{\psi_E} \lra\\
\sum_n (E+E_0-\ep_n) \ket{n} \braket{n}{\psi_E} = \sum_q (j_q+E_0)\ket{q}\braket{q}{\psi_E}
\lra \\
\sum_p \frac{\ket{q}\bra{q}}{j_q+E_0}\sum_n (E+E_0-\ep_n) \ket{n} \braket{n}{\psi_E} = \ket{\psi_E} \lra \\
\lrb{\sum_n \ep_n \ket{n}\bra{n} + \sum_{n,m}j_{mn}^{\rm eff}\ket{m}\bra{n}}\ket{\psi_E} = E\ket{\psi_E}  \ ,
\end{multline}
with
\be
j_{mn}^{\rm eff} = (E+E_0-\ep_m) \sum_q \frac{\braket{m}{q}\braket{q}{n}}{j_q+E_0}(E+E_0-\ep_n) - \delta_{m,n} E_0 \ .
\ee
In such a way one can immediately show localization of the bulk eigenstates as the above effective hopping term for $a<1$ decays as
\be\label{eq:j_eff_a<1_basic}
j_{m,m+R}^{\rm eff} \sim \frac{W^2}{R^{2-a} t}
\ee
together with the eigenstate tails confirming the duality suggested in~\cite{Deng2018duality}.
In the further sections we generalize this method to $t\sim N^{\gamma/2}$.

\subsection{Spatial renormalization group}\label{Sec:RG}
In this section, we consider a renormalization group (RG) derived for the BM model in~\cite{Burin1989} and developed further in~\cite{Kutlin2020_PLE-RG} beyond the strong-disorder limit.
The main idea of the RG is as follows. First, one cuts off the tunneling beyond a certain scale $R_0$ and calculate the wave functions ($R_0$ modes) for this scale.
Then one chooses a new cutoff $R>R_0$ and constructs new modes ($R$ modes) a superposition of $R_0$ modes, taking into account the resonances between $R_0$ modes.
The localization length increases from $l_0 \lesssim R_0$ to $l_1 \lesssim R$ due to the presence of these resonances.
Due to the smallness of the parameter $j_{\rm eff}/W\ll1$ only pairs of resonances are taken into account.
The $R$ modes $\ket{\psi_k^{(1)}}$ can be written via the initial site basis vectors $\ket{n}$ as follows
\be
\ket{\psi_k^{(1)}} = \sum_n \psi_k^{(1)}(n)\ket{n} \ .
\ee
Thus, the hopping term rewritten in the new operators takes the form
\be\label{eq:j_{mn}_psi}
\sum_{m,n} j_{mn} \ket{m}\bra{n} = \sum_{k,l}\ket{\psi_k^{(1)}}\bra{\psi_l^{(1)}}\sum_{m,n}\psi_k^{(1)}(m)\psi_l^{(1)*}(n) j_{mn} \ .
\ee
According to RG assumption the modes $\psi_k^{(1)}(m)$ are localized within the interval $|k-m|<l_1$ at the length $l_1\lesssim R$ for the deterministic potential
$j_{mn} = t/||m-n||^a$. Therefore one can neglect the difference between $j_{mn}$ and $j_{kl}$ ($\lv|i-j|-|k-l|\rv<|i-k|+|j-l|<2 l_1\lesssim R$) due to the smoothness of the potential (see~\cite{Kutlin2020_PLE-RG} for more rigorous consideration).
As a result, Eq.~\eqref{eq:j_{mn}_psi} reads as
\be
\sum_{m,n} \frac{t}{||m-n||^a} \ket{m}\bra{n} \simeq \sum_{k,l}\frac{t \ell_k \ell_l^*}{||k-l||^a}\ket{\psi_k^{(1)}}\bra{\psi_l^{(1)}}, \quad \ell_k = \sum_{n}\psi_k^{(1)}(n) \ .
\ee

In order to estimate the renormalized hopping term ${t \ell_k \ell_l^*}/{||k-l||^a}$
one should consider the typical amplitude of $\ell_k$ at a certain energy $E$ as follows
\begin{multline}
\la \ell^2 \ra_E = \frac{\la \sum\limits_k \ell_k^2 \delta (E-E_k)\ra}{g(E)} = \frac{\la \sum\limits_k \sum\limits_{m,n\atop |m-k|,|n-k|<R}\psi_k^{(1)}(m)\psi_k^{(1)*}(n) \delta (E-E_k)\ra}{g(E)} \simeq \\\frac{\sum\limits_{|m-n|<R} \Im {\bar G}_{mn}}{g(E)} \simeq \frac{\Im {\bar G}_{q\simeq 1/R}(E)}{g(E)},
\end{multline}
where the global density of states (DOS) is given by
\be\label{eq:g(E)_sum}
g(E) = \la \sum_k \delta (E-E_k)\ra = \frac{\Im {\bar G}_{mm}(E)}{\pi} , \quad {\bar G}_{mm}(E) = \bracket{m}{\hat {\bar G}(E)}{m} = \sum_q \frac{{\bar G}_{q}(E)}{N},
\ee
$G_{m-n}$ and $G_q$ stand for the Green's function in the real and momentum space satisfying the equation
$
(E-\hat H)\hat G(E) =  \hat 1
$, 
and the bar on top it stands for the disorder averaging over the diagonal elements $\ep_n$.

Using the standard coherent potential approximation (see, e.g.,~\cite{Yonezawa1973coherent_potential}), for completeness considered in Appendix~\ref{AppSec:Coherent_potential},
one can find both the averaged Green's function, diagonal in the momentum space, and its self-energy part $\Sigma$ given by the following self-consistency equations
\be\label{eq:G(E)_Sigma}
\hat{\bar G}(E) = \sum_q\frac{\ket{q}\bra{q}}{E-\Sigma -j_{q}}  \ , \quad
\Sigma = \int \frac{\ep P(\ep)d\ep}{1 - (\ep - \Sigma) {\bar G}_{mm}(E) } \ ,
\ee
where $P(\ep)$ is the distribution of the diagonal disorder $\ep_n$.

Note that the above results can be checked numerically by calculating the eigenstate spatial decay from the maximum $k=n$
\be\label{eq:psi_typ_decay}
|\psi_{typ}|^2(R) = \exp\lrb{\mean{\ln|\psi_n(k=n+R)|^2}} \sim \lrp{\frac{j_R^{\rm eff}}{W}}^2 \ .
\ee
The calculation of the moment of the wave function $\mean{|\psi|^{q}}(R)$ has a divergence over the energy difference $\ep_n - \ep_k$ in the denominator and, thus, one can use the Breit-Wigner approximation in order to calculate it (see, e.g.,~\cite{Kutlin2020_PLE-RG}):
\be
|\psi_{n}(k=n+R)|^2 \simeq \frac{\lrp{j_R^{\rm eff}}^2 }{\lrp{\ep_n - \ep_k}^2+\lrp{j_R^{\rm eff}}^2}
\ee
and, thus,
\be
\mean{|\psi|^{q}}(R) \simeq \left\{
\begin{array}{ll}
\lrp{j_R^{\rm eff}}^q & q<1 \\
j_R^{\rm eff}\ln\lrp{\frac{2 W}{j_R^{\rm eff}}} & q = 1 \\
j_R^{\rm eff} & q>1 \\
\end{array}
\right.
\ee

From this, one can easily calculate the spectrum of fractal dimension $f(\alpha)$ via the number of points at the distance $R$
\be\label{eq:f(alpha)_def}
N^{f(\alpha_R)}\sim R
\ee
where the typical wave function scales as $|\psi_{typ}|(R)^2\sim N^{-\alpha_R}$.


With this general description, in the further sections we focus on the Burin-Maksimov model~\cite{Burin1989} in 1d~\cite{Deng2018duality,Nosov2019correlation}
generalized to the $N$-dependent prefactor of the hopping term.

\section{Numerical results}\label{Sec:Results}

Starting with the numerical simulations, we show the eigenstate spatial structure, including the short-range behavior and the long-range power-law decay~\eqref{eq:psi_typ_decay}.
Then we explicitly determine the spectrum of fractal dimensions and show it for mid-spectrum states.
In order to demonstrate the similarity to the MBL problem in the end of the section we also show the adjacent gap ratio statistics of eigenvalues.

\begin{figure}[tb]
	\center{
	\includegraphics[width=0.49\textwidth]{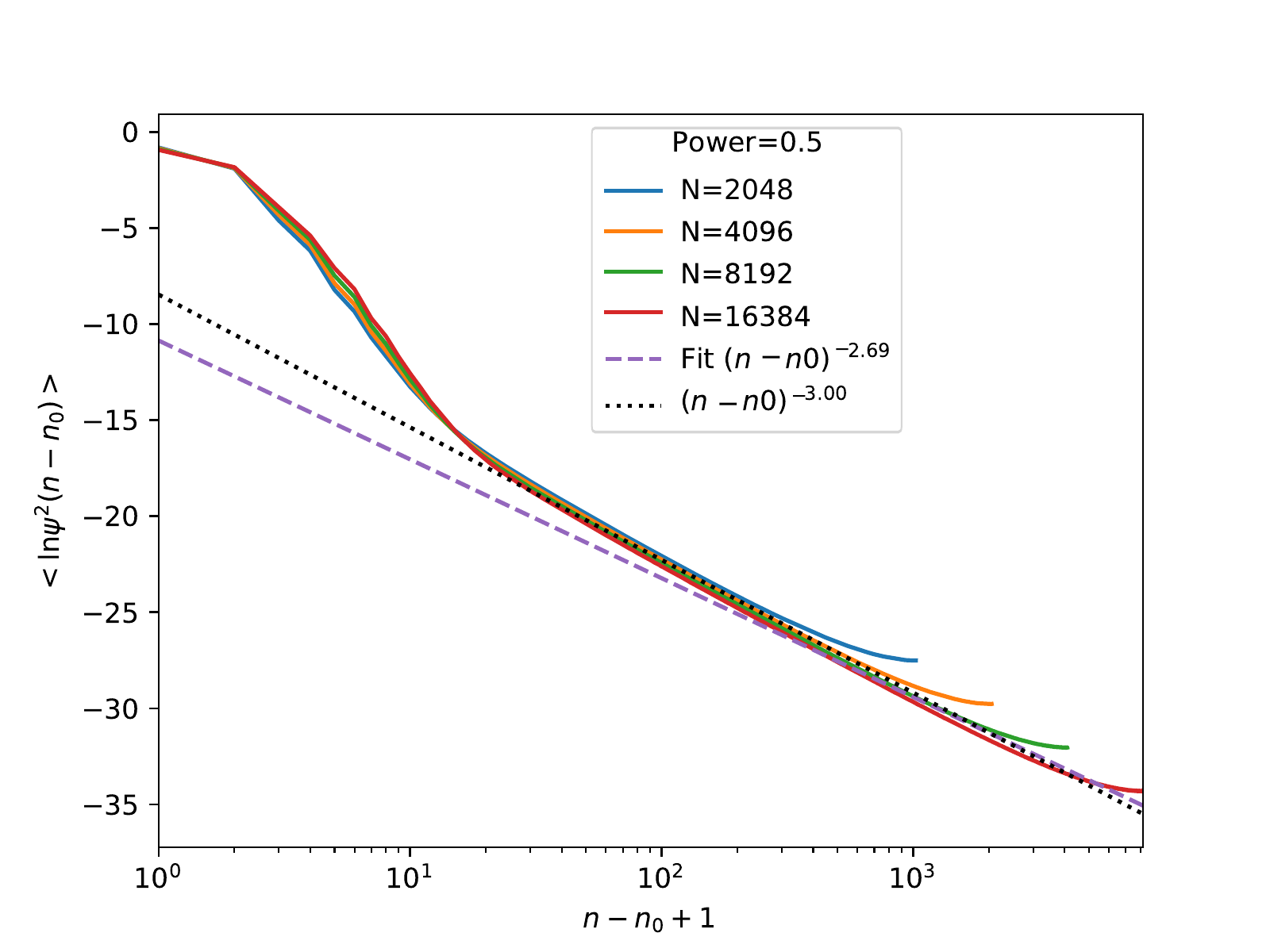}
	\includegraphics[width=0.49\textwidth]{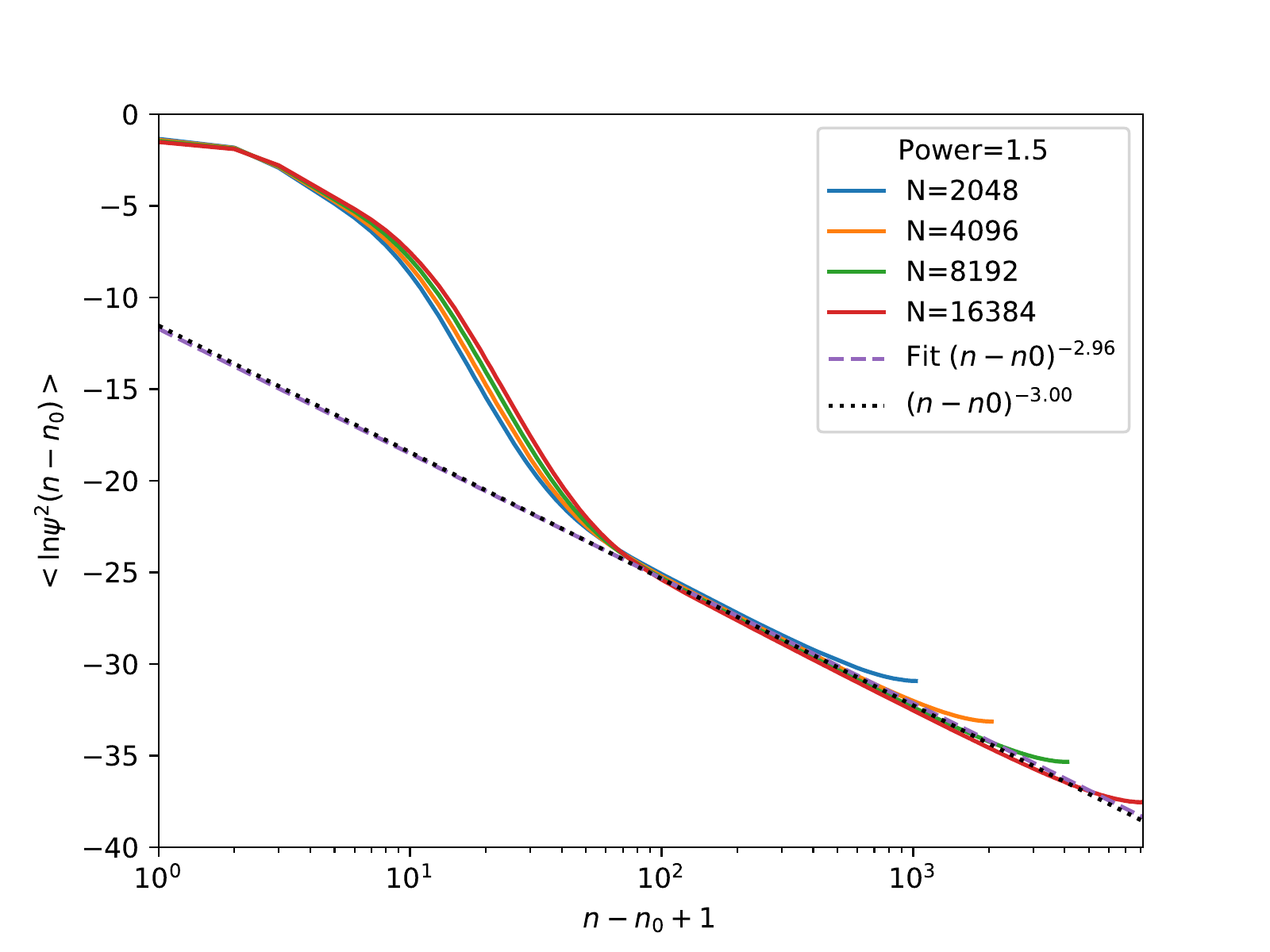}
	\caption{{\bf The bulk-spectrum eigenstate spatial decay,~\eqref{eq:psi_typ_decay}, at large distances $R=||n-n_0||\gg R_*$ in log-log scale} for $\gamma=0.25$ and (left)~$a=0.5$, (right)~$a=1.5$.
    The data for the system sizes from $N=2^{12}$ to $2^{14}$ taken within $10$\% of the spectrum around DOS maximum is averaged over $400$ disorder realizations and accompanied by the power-law fit (violet dashed line) and by the theoretical prediction (black dotted line) which show good agreement.
	\label{fig:PL-decay}
    }
    }
\end{figure}

Figure~\ref{fig:PL-decay} shows the long-distance power-law decay, Eq.~\eqref{eq:psi_typ_decay}, from the eigenstate maximum
\be\label{eq:psi_decay}
|\psi_E(R)|\sim 1/R^{a_{\rm eff}} \ , \quad a_{\rm eff} = \max\lrp{a,2-a} \ .
\ee
In order to emphasize the power-law nature of the decay both panels are shown in log-log scale.
Like in the standard BM model~\cite{Nosov2019correlation} or in its Euclidean version~\cite{Kutlin2020_PLE-RG} the exponent of this power-law decay is symmetric with respect to the critical point $a=d=1$. However, unlike the above models one can see that the short-range behavior of the wave functions depends on the system size.

\begin{figure}[tb]
	\center{
	\includegraphics[width=0.49\textwidth]{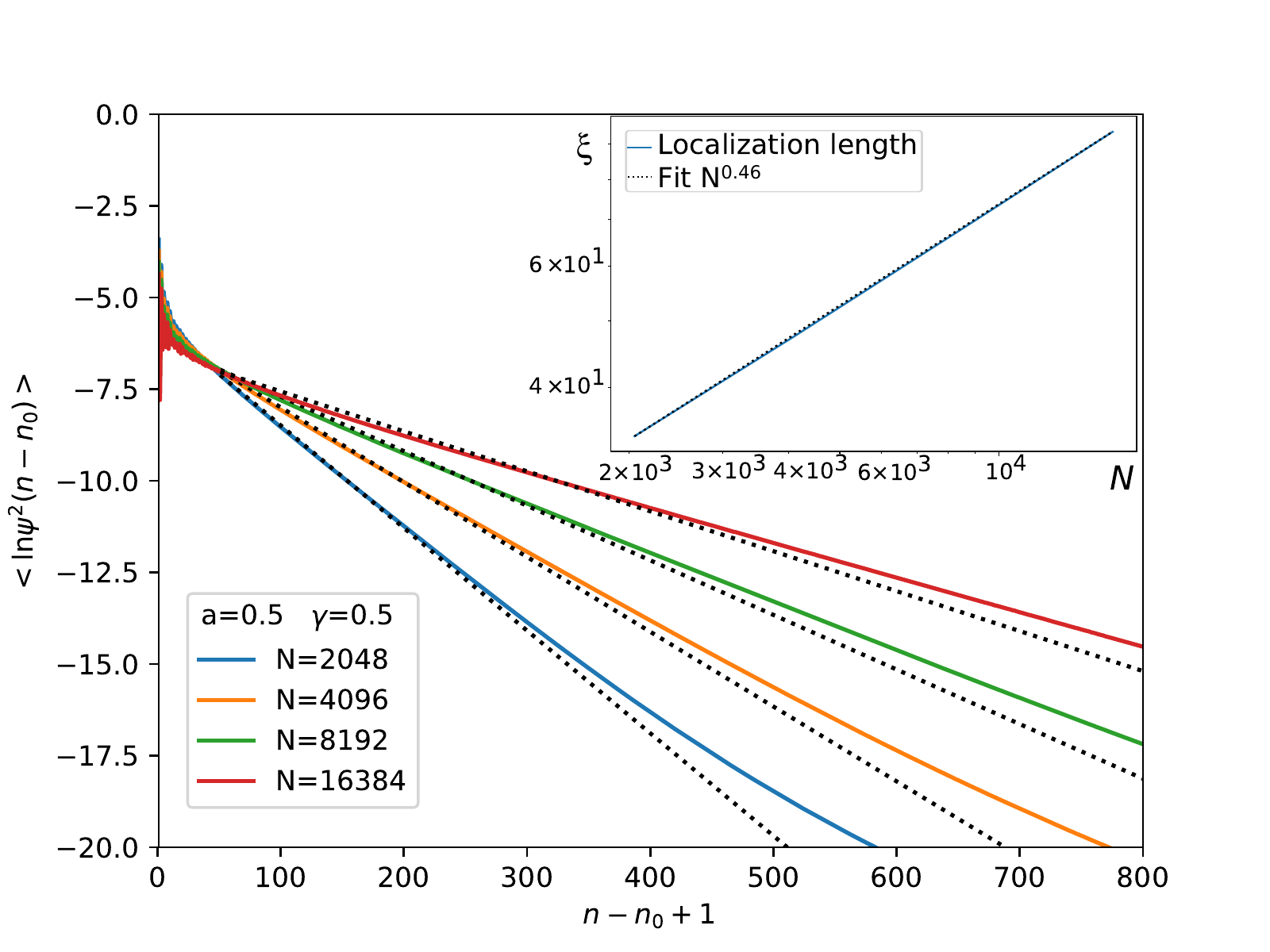}
	\includegraphics[width=0.49\textwidth]{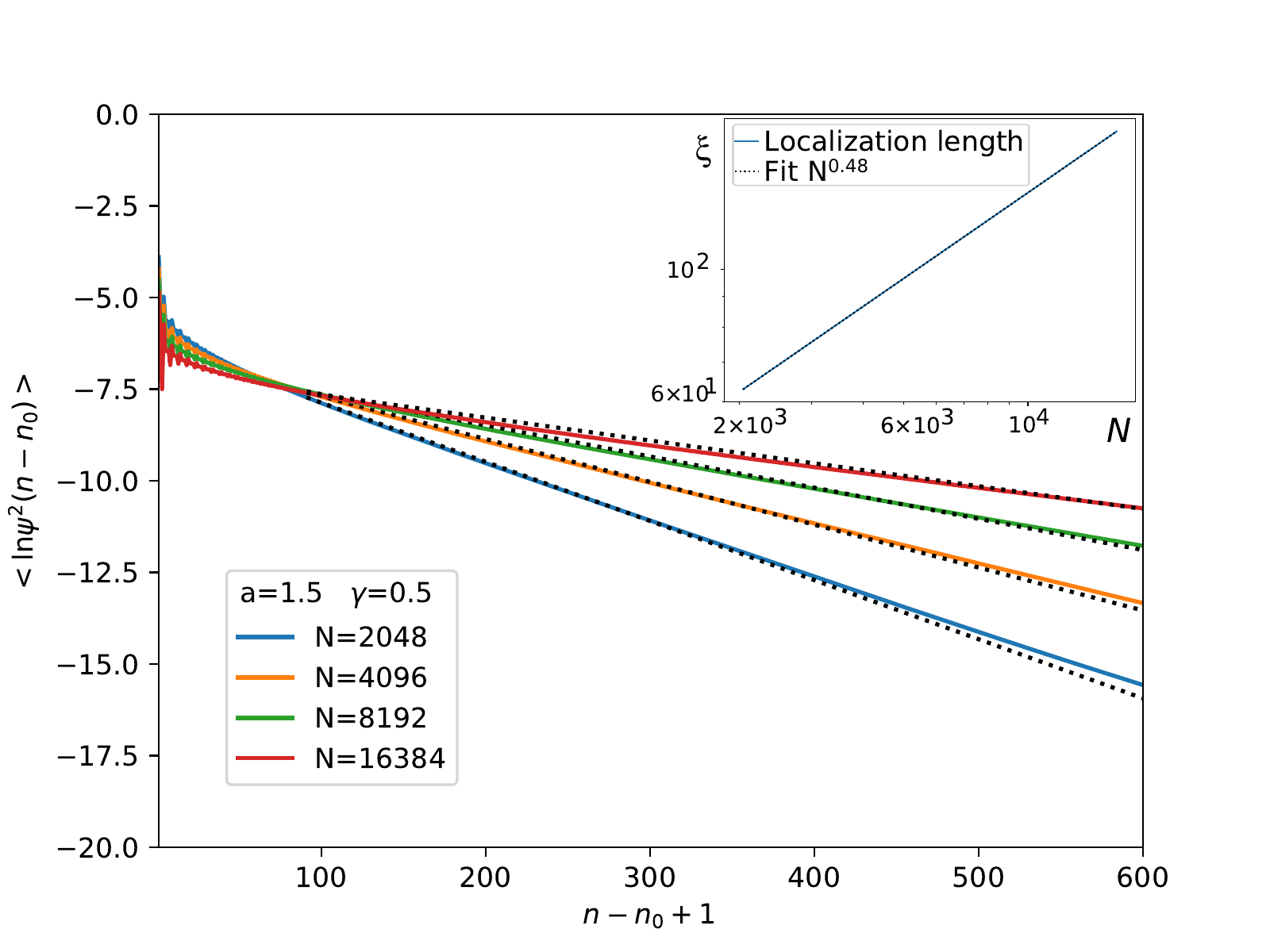}\\
	\includegraphics[width=0.49\textwidth]{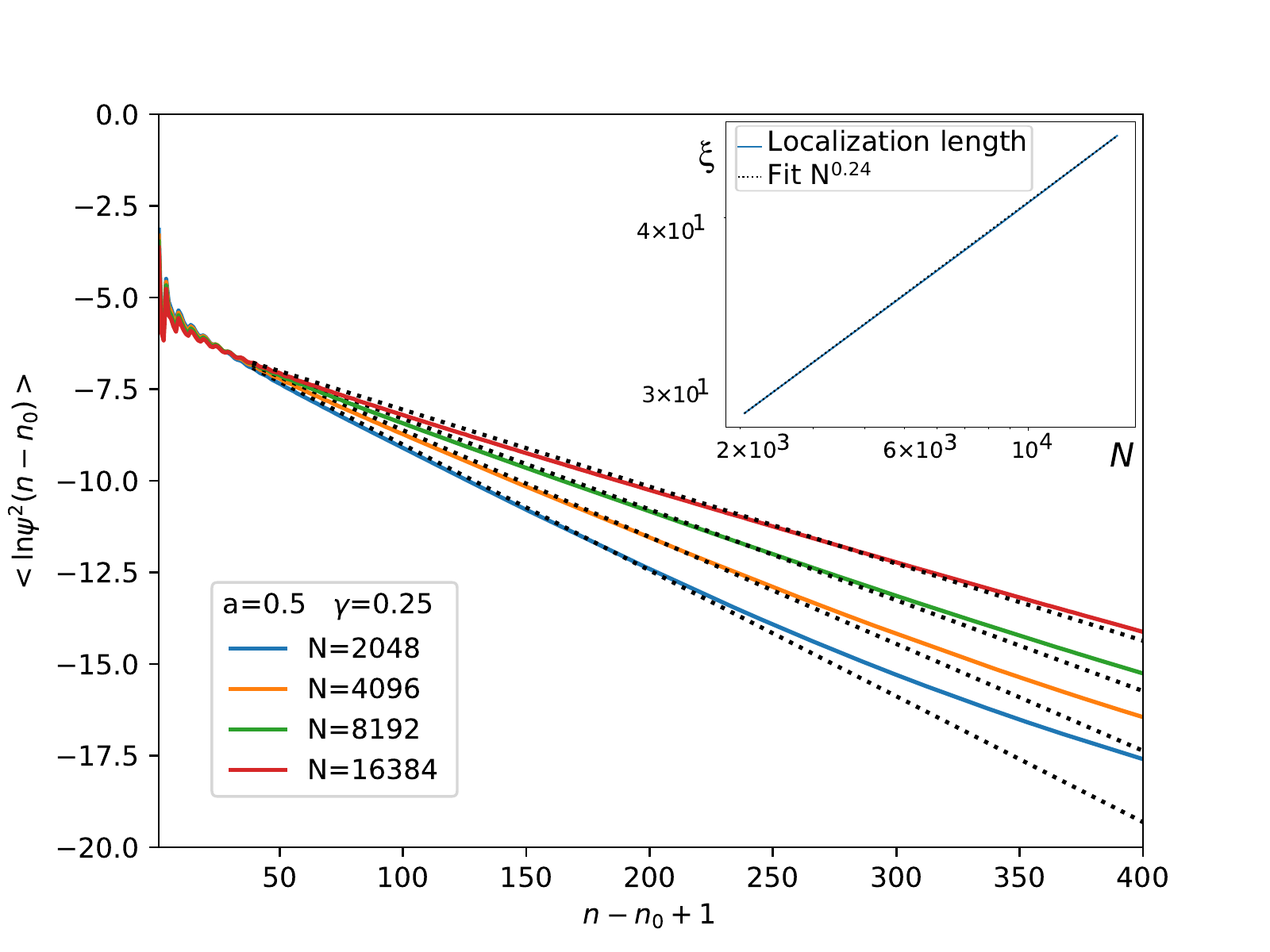}
	\includegraphics[width=0.49\textwidth]{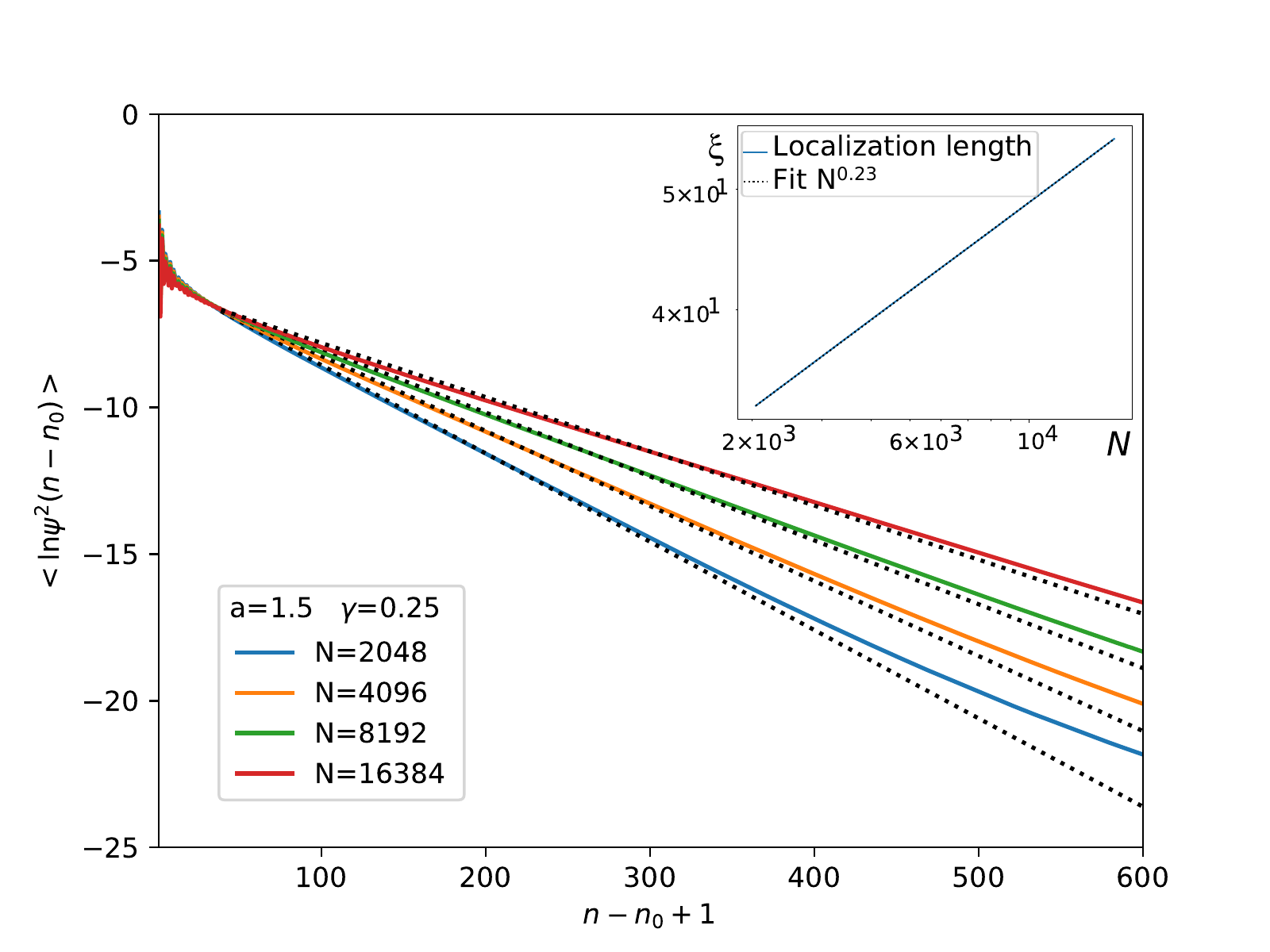}
	\caption{{\bf The bulk-spectrum eigenstate spatial decay at small distances $R=||n-n_0||\ll R_*$ in log-linear scale} for
    (left~panels)~$a=0.5$,
    (right~panels)~$a=1.5$ and
    (top~panels)~$\gamma=0.5$,
    (bottom~panels)~$\gamma=0.25$.
    The data for the system sizes from $N=2^{12}$ to $2^{14}$ averaged over $400$ disorder realizations is taken within $20$\% of the spectrum in the middle and accompanied by the exponential fit (black dotted line) with the inverse decay rate (effective localization length) $\xi$ scaling with the system size.
    The insets show the power-law system-size scaling of $\xi$, which shows good agreement with the theoretical prediction $\xi\sim N^{\gamma}$, Eq.~\eqref{eq:loc_length_bulk}.
	\label{fig:exp-decay}
    }
    }
\end{figure}

In order to clarify the system-size dependence of the short-range eigenstate decay, in Fig.~\ref{fig:exp-decay} we show the zoom of this distance range for several power-law exponents $a=0.5$ and $1.5$ and hopping scaling $\gamma=0.25$ and $0.5$. From the log-linear plots one can immediately recognize an exponential decay of eigenstates, similar to the one in the short-range Anderson models, $|\psi_E(R)|\sim e^{-R/\xi}$.
In addition, one can see that the effective localization length $\xi$ of this exponential decay grows with the system size scaling as a power-law
\be\label{eq:xi_N^gamma}
\xi \sim N^{\gamma} \ .
\ee
This scaling is extracted from the exponential fits of the data in the main panels and shown in the corresponding insets.

\begin{figure}[t!]
	\center{
	\includegraphics[width=0.4\textwidth]{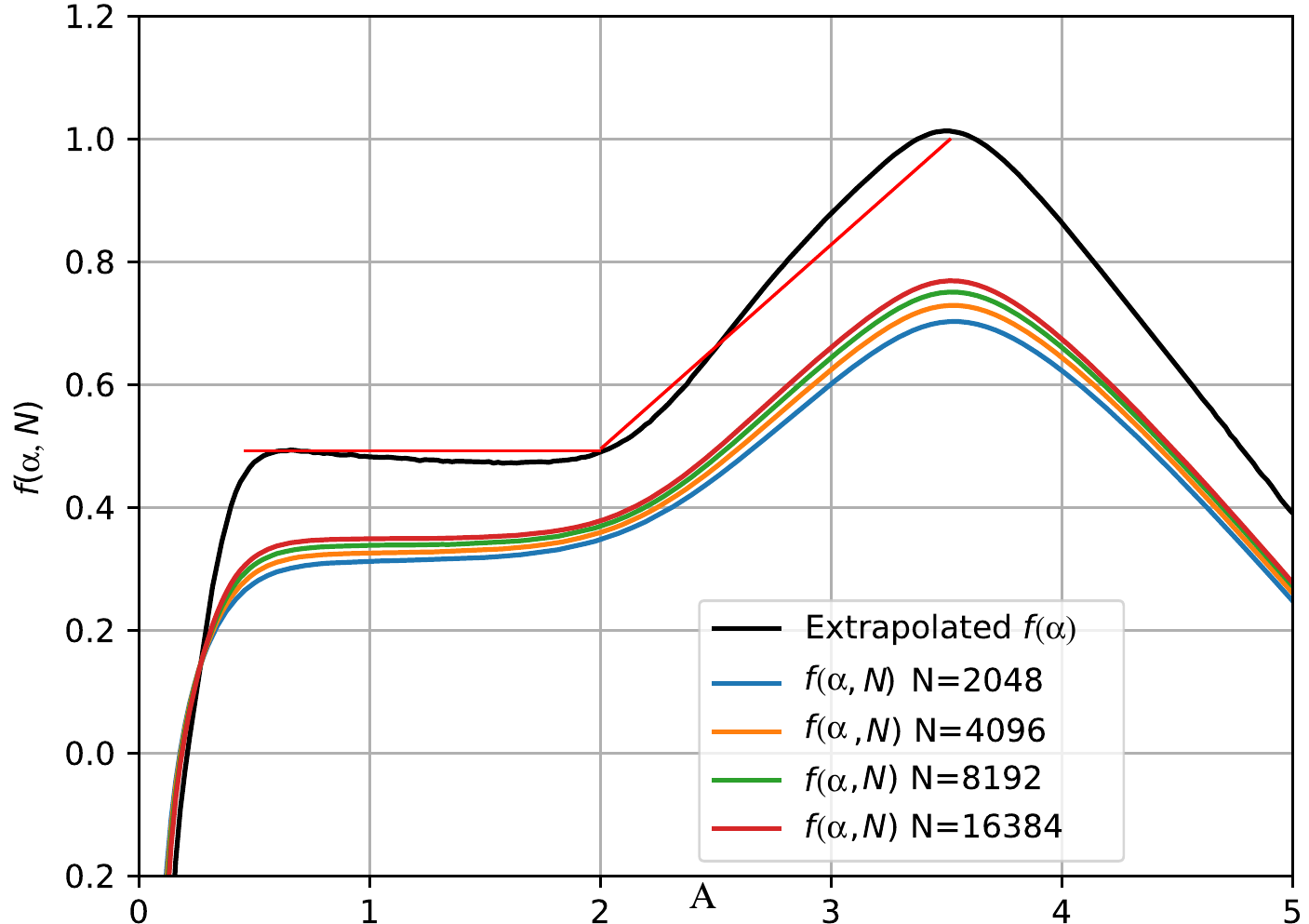}
	\includegraphics[width=0.4\textwidth]{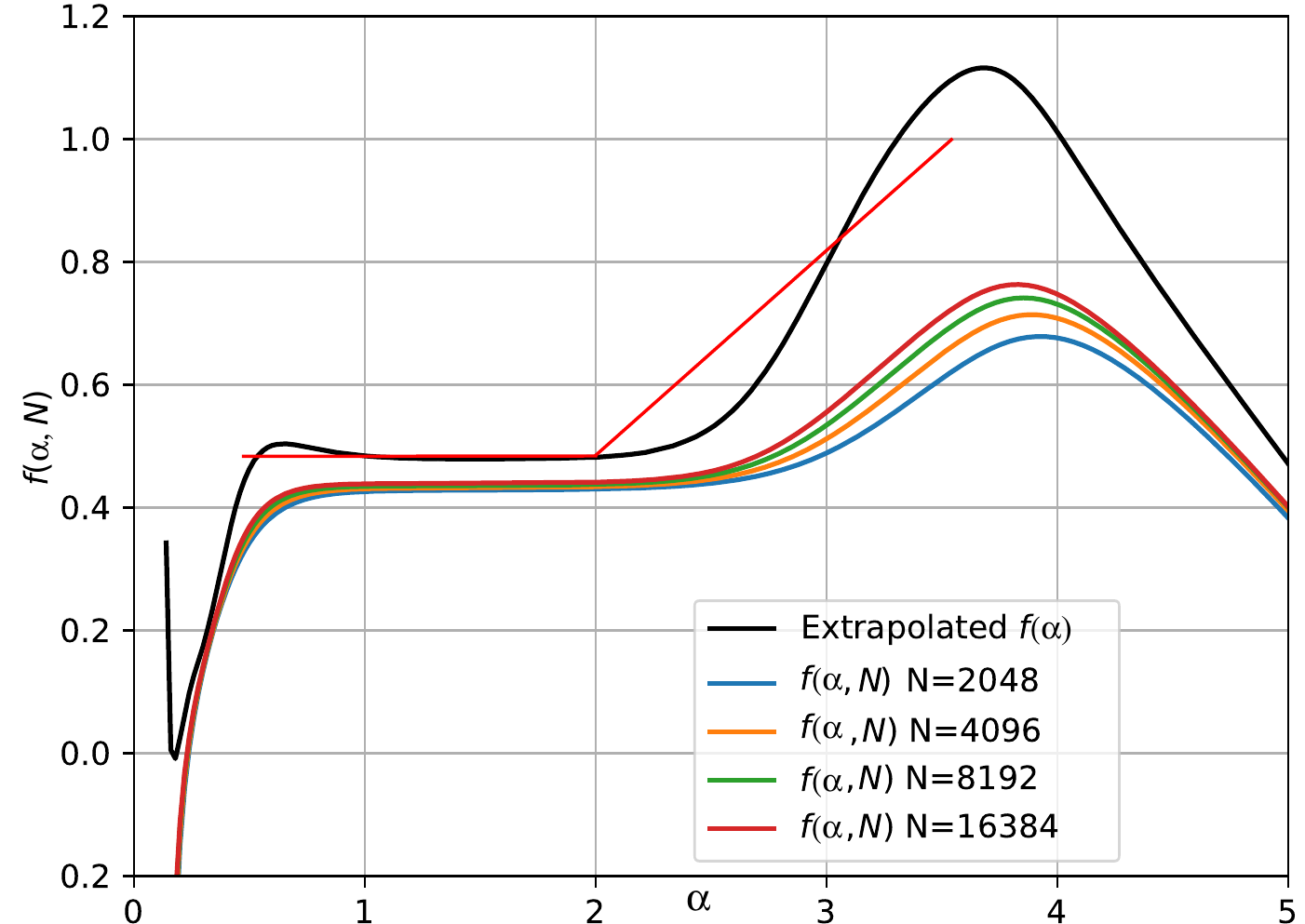}
	\caption{{\bf Spectrum of fractal dimensions $f(\alpha)$ of the bulk-spectrum eigenstates} for $\gamma=0.5$, (left)~$a=0.5$ and $25$\%, (right)~$a=1.5$ and $15$\% of the spectrum around the DOS maximum.
    The data averaged over $400$ disorder realizations is extrapolated from the system sizes $N=2^{12}$~-~$2^{14}$ to infinity using the standard extrapolation techniques (see, e.g.,~\cite{Nosov2019correlation}).
	\label{fig:f(a)}
    }
    }
\end{figure}

Next, we consider the spectrum of fractal dimensions, which is defined as the power $f(\alpha)$ of the scaling of the distribution $P(\alpha)\sim N^{f(\alpha)-1}$ of $\alpha=-\ln |\psi_E(n)|^2/\ln N\geq 0$~\cite{Evers2008anderson}
\be\label{eq:f(a)}
f(\alpha) = 1+\lim_{N\to\infty} \frac{\ln\lrb{P\lrp{\alpha=-2\ln|\psi_E(n)|/\ln N}}}{\ln N} \ .
\ee
This function $f(\alpha)$ characterizes the fractal dimensions $D_q$ via the wave-function moments averaged over disorder and via the corresponding Legendre transform
\be
\mean{\sum_n \abs{\psi_E(n)}^{2q}}\sim N^{(1-q)D_q} = \int N^{-q\alpha}P(\alpha) d\alpha \lra (q-1) D_q = \min_\alpha \lrb{q \alpha - f(\alpha)} \ .
\ee
Among many properties of $f(\alpha)$ (like the normalization conditions of the probability distribution, and of the wave function) we would like to focus on the so-called multifractal symmetry~\cite{Evers2008anderson}
\be\label{eq:f(a)_symm}
f(2-\alpha) = f(\alpha) - (\alpha-1) \ ,
\ee
which relates wave-function tails (large $\alpha>1$) with peaks (small $\alpha<1$).
This symmetry is known to work for the extended states, fractal or multifractal with some (at least critical) level repulsion.
However as we show in Fig.~\ref{fig:f(a)}, the spectrum of fractal dimension in the bulk spectral states of the generalized BM model,~\eqref{eq:ham_mat}~-~\eqref{eq:dist},
violates this symmetry by combining an approximate plateau $f(\alpha) \simeq \gamma$ at $\alpha\gtrsim \gamma$ corresponding to the exponential decay shown in Fig.~\ref{fig:exp-decay}
with the linear ramp $f(\alpha) = (\alpha-\gamma)/(2 a_{\rm eff})$ governed by the power-law decay~\eqref{eq:psi_decay}.
The violation of the symmetry~\eqref{eq:f(a)_symm} implicitly confirms the Poisson level statistics.

\begin{figure}[h!]
	\center{
	\includegraphics[width=0.4\textwidth]{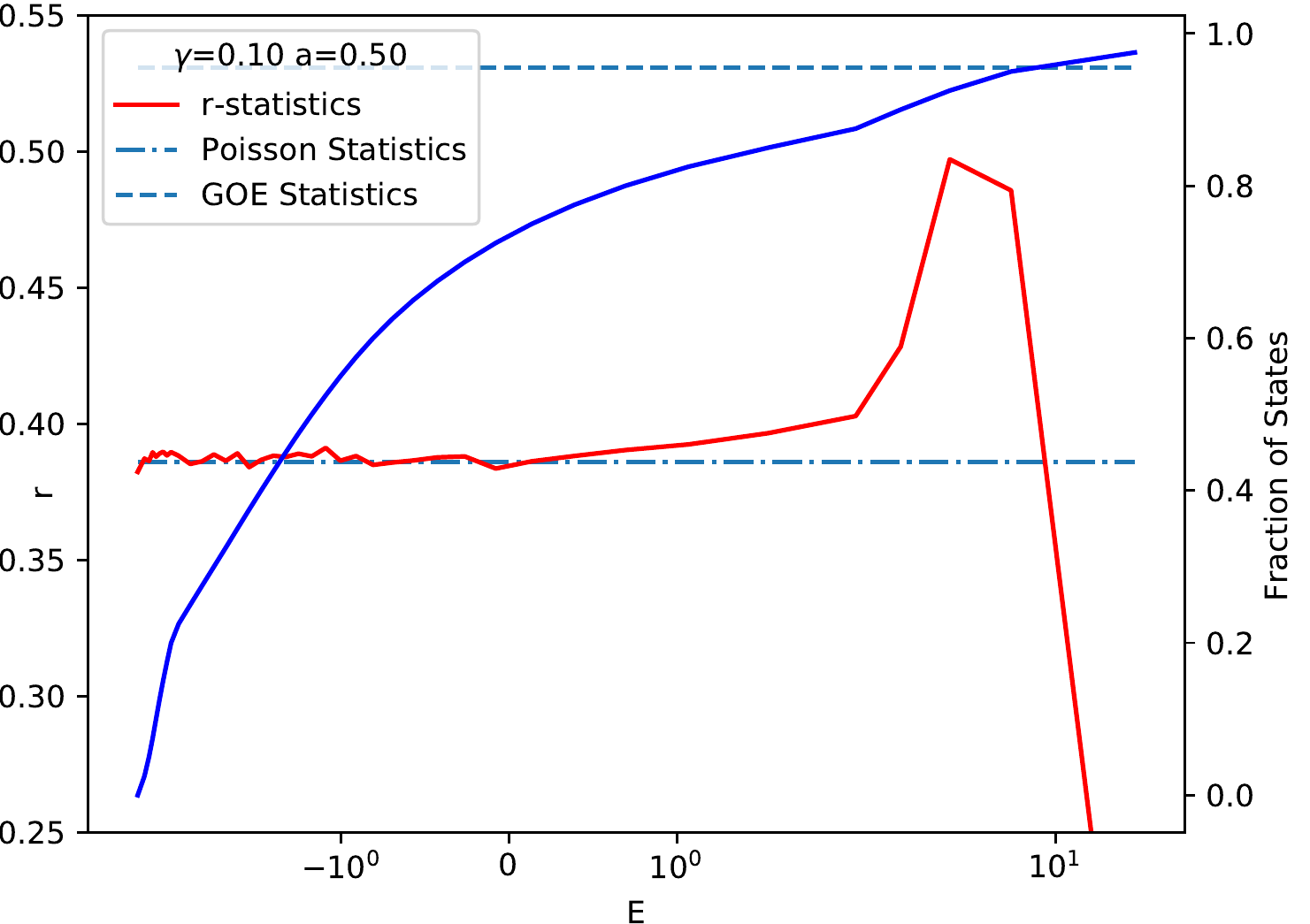}
	\includegraphics[width=0.4\textwidth]{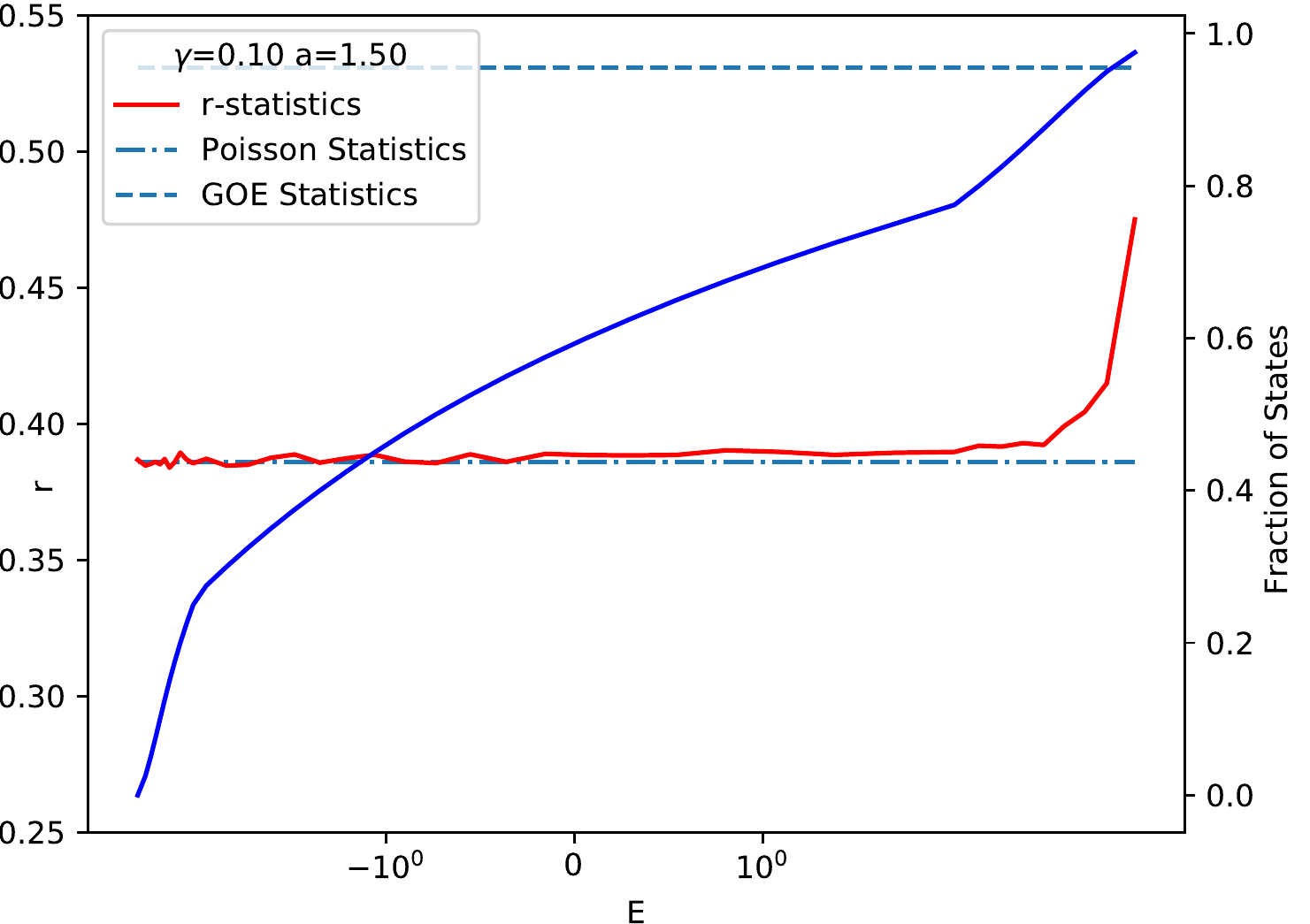}
	\caption{{\bf The adjacent gap ratio statistics} ($r$-statistics) across the spectrum of the generalized Burin-Maksimov model,~\eqref{eq:ham_mat}~-~\eqref{eq:dist}, for $\gamma=0.1$ and (left)~$a=0.5$, (right)~$a=1.5$. The $r$-statistics (red) averaged over $100$ disorder realizations for the system size $N=2^{14}$ is accompanied by the fraction of states (blue) or, in other words, the cumulative distribution function showing that in both cases at least $80$\% of the states (at this system size) demonstrate Poisson level statistics.
	\label{fig:r-stat}
    }
    }
\end{figure}

In order to demonstrate the Poisson eigenlevel statistics explicitly, in Fig.~\ref{fig:r-stat} we show the adjacent gap ratio statistics introduced in~\cite{oganesyan2007localization,Atas2013distribution} as
\be\label{eq:r-stat}
r_n = \frac{\min\lrp{s_n, s_{n+1}}}{\max\lrp{s_n, s_{n+1}}} \text{, where } s_n = E_{n+1} - E_n \ .
\ee
Here we will consider this $r$-statistics averaged over disorder realizations, $r=\mean{r_n}$.
The absence of the level repulsion, i.e. Poisson statistics, corresponds to $r=r_P = 2\ln 2 - 1 \simeq 0.3863$, while
the Wigner-Dyson level repulsion~\cite{Mehta2004random} leads to $r=r_{GOE} \simeq 0.5307$ for the orthogonal symmetry to which we stick here~\footnote{Some models show
Wigner-Dyson statistics for non-ergodic extended eigenstates. In such situation people has to use either higher order gap ratios (like in~\cite{Atas2013joint,Tekur2018higher,deng2020anisotropy}) or other global spectral measures like spectral form factor, level compressibility~\cite{Kravtsov_NJP2015}, and power spectrum (see below).}.
Figure~\ref{fig:r-stat} shows $r$-statistics, averaged over disorder, versus the energy.
As the spectrum in BM model is highly inhomogeneous~\eqref{eq:j_q}, in addition we show the corresponding fraction of states at each energy, given by the integration of DOS and called a cumulative distribution function.
One sees the Poisson level statistics, $r=r_P$, in the extensive interval of energies corresponding to most of the states, except a small fraction (which is of the order of $20$\% for the considered system size) of the topmost energy states, corresponding to the momentum-space localization,~\eqref{eq:q^*}~\footnote{In the case of $a<1$, the rapid drop of the $r$-statistic at the high energy range is due to those momentum-space localized states, which are nearly degenerate due to the momentum inversion symmetry.}.

\begin{figure}[h!]
	\center{
	\includegraphics[width=0.8\textwidth]{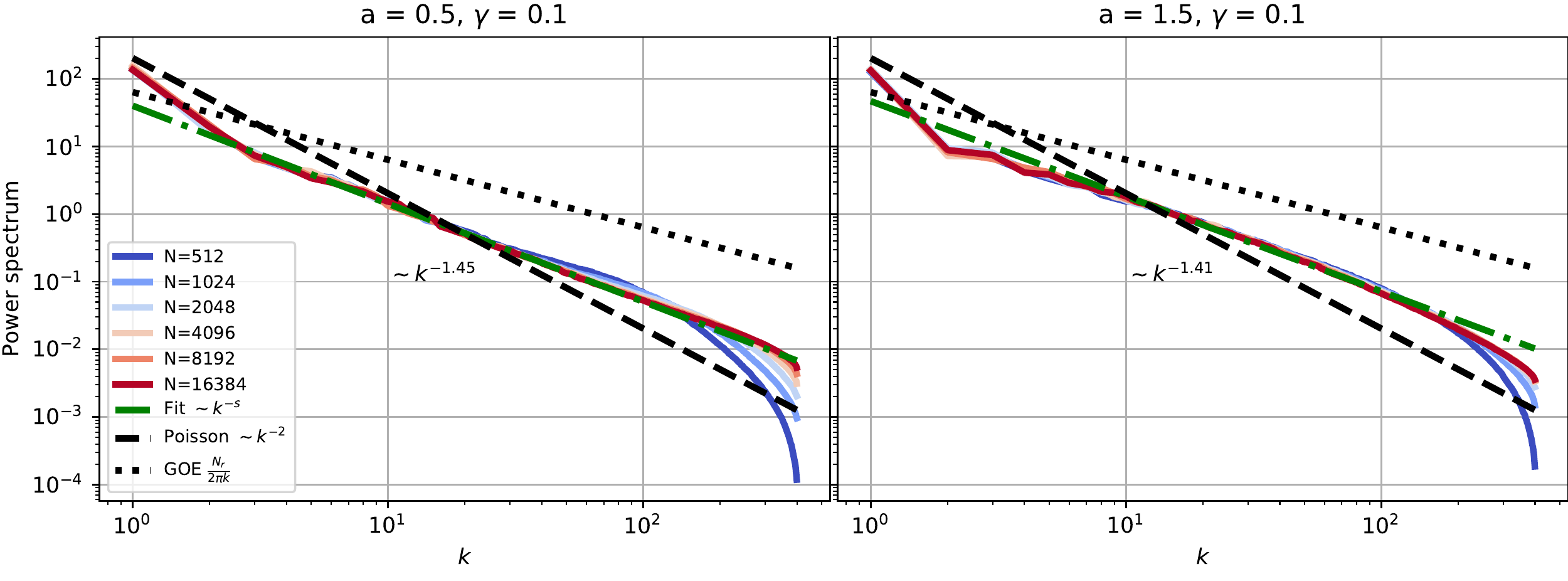}\\
	\includegraphics[width=0.8\textwidth]{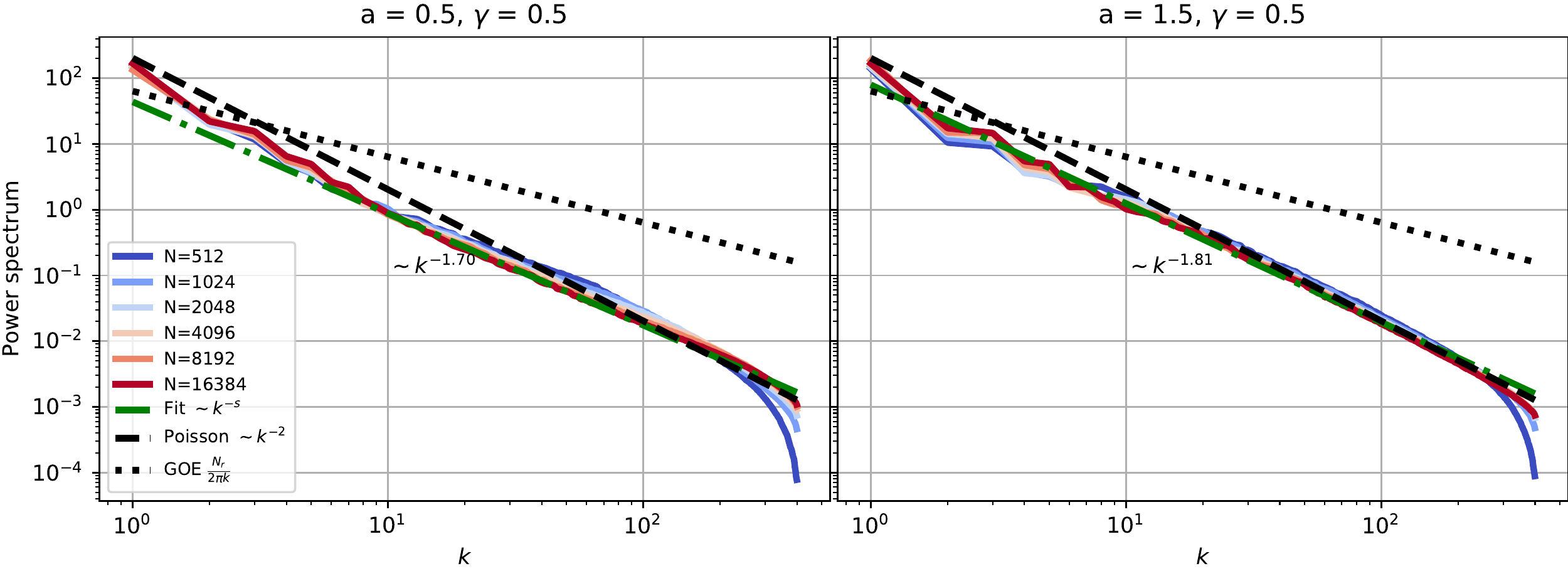}
	\caption{{\bf The power spectrum} ($PS$), defined in Eq.~\eqref{eq:PS_def}, is shown for the generalized Burin-Maksimov model,~\eqref{eq:ham_mat}~-~\eqref{eq:dist}, for (top)~$\gamma=0.1$, (bottom)~$\gamma=0.5$ and (left)~$a=0.5$, (right)~$a=1.5$ for different system sizes (see legend).
The theoretical curves for Wigner-Dyson $PS = N_r/\lrp{2\pi k}$ (dotted) and Poisson $PS \sim k^{-2}$ (dashed) are shown by black lines.
The power spectrum averaged over $400$ disorder realizations for each system size.
The curve for $N=2^{14}$ is accompanied by the power-law fit $PS\sim k^{-s}$ in the interval $5\leq k\leq 100$, shown as a green dash-dotted line.
The corresponding power $s$ is written close to the curve in each panel.
	\label{fig:PS}
    }
    }
\end{figure}

In order to uncover global spectral correlations, we have also considered a so-called power spectrum~\cite{Riser2017power,Riser2020nonperturbative,Riser2021power,Riser2021power,Berkovits2020super,Berkovits2021probing,colmenarez2021spectral}, see Fig.~\ref{fig:PS}.
Unlike $r$-statistics, this measure, similar to the level rigidity, is a global spectral characteristics, which can distinguish not only Wigner-Dyson and Poisson level statistics, but also can pinpoint the energy scale of the crossover between them~\cite{Berkovits2020super,Berkovits2021probing,colmenarez2021spectral}.
In the simplest incarnation~\cite{Berkovits2020super}, which does not need any unfolding, the power spectrum $PS_k$ is the intensity $\lambda_k^2$ of the eigenvalues of a singular value decomposition of the rectangular matrix $\mathcal{E}$, containing $N_R$ realizations, $\mathcal{E}_{k,n}$, $1\leq k\leq N_r$, of the spectrum, with $N$ eigenvalues in each, $1\leq n\leq N$:
\be\label{eq:PS_def}
\mathcal{E} = U \Lambda V \ , \quad \Lambda = {\rm diag}\lrp{\lambda_1,\ldots,\lambda_{N_r}} \lra PS_k = \lambda_k^2 \ .
\ee
Here unitary matrices $U$ and $V$ are $N_r\times N_r$ and $N\times N$, respectively, and
we consider $N_r\leq N$ and order eigenvalues $\lambda_k$ in the non-increasing order $\lambda_k\geq\lambda_{k+1}$.
Note that the singular values at $k\ll N_r$ capture the global trends of the spectra,
while the ones at $k\sim N_r$ represent the local fluctuations (similar to the $r$-statistics).

In such a definition, the Wigner-Dyson level statistics corresponds to the $1/f$-``noise'' behavior of the power spectrum~\cite{TorresVargas2017determination,TorresVargas2018crossover} (dotted black line in Fig.~\ref{fig:PS}, $f = 2\pi k/N_r$,
while the pure Poisson level statistics gives $PS\sim 1/f^2$ (dashed black line).

From the Fig.~\ref{fig:PS}, one can see that for all considered $0<\gamma<1$ the the high-$k$ behavior of the power spectrum coinsides indeed to the Poisson one in agreement with the $r$-statistics, while the low-$k$ $PS$-values show the tendency of even faster decay.
The finite-size power-law exponent (extracted from a power-law fit, green line) is significantly larger than $1$ and flows towards $2$ with increasing system size.
All this confirms the Poisson level statistics in the bulk of the spectrum, while the low-$k$ contribution is related to the
degeneracy of the high-energy momentum-space localized states.

\section{Analytical consideration}\label{Sec:analytics}
In this section, we provide the analytical consideration of the generalized BM model,~\eqref{eq:ham_mat}~-~\eqref{eq:dist}, based on the renormalization group analysis from Sec.~\ref{Sec:RG} supplemented by the coherent potential approximation, Appendix~\ref{AppSec:Coherent_potential}.

First, we consider the long-range asymptotic power-law decay of the eigenstates~\eqref{eq:psi_typ_decay}.
For this purpose, we take into account the Lorenzian form~\eqref{eq:G(E)_Sigma} of the imaginary part of the Green's function in the momentum basis
\be\label{eq:ImG_Lorenz}
\Im {\bar G}_q(E) \simeq \frac{\Gamma}{(E - j_q)^2+\Gamma^2} \ ,
\ee
where for simplicity we consider the box distribution of the diagonal disorder $\ep_n$ and focus on the limit in the spectral bulk, where $\bar{G}_{mm}(E)\ll 1/W$,
i.e. the level broadening, $\Gamma = \Im \Sigma$, being the imaginary part of the self-energy $\Sigma$, is given by Eq.~\eqref{eq:Gamma_limits}, $\Gamma=\pi g(E) {W^2}/{12}$.
In this case, one can calculate the effective hopping matrix element as follows
\be\label{eq:j_eff}
j_{R}^{\rm eff} \equiv \frac{t l_k l_{k+R}^*}{R^a} \lesssim
\frac{t}{R^a}\frac{\Im {\bar G}_{q=1/R}(E)}{g(E)} 
\ee
with the DOS given by~\eqref{eq:g(E)_sum}:
\be\label{eq:DOS_sum}
g(E) = \frac{\Im {\bar G}_{0}(E)}{\pi} \simeq \frac{1}{\pi N}\sum_q \frac{\Gamma}{(E - j_q)^2 + \Gamma^2} \simeq \min\lrp{\frac1W, \abs{\frac{d q}{d j_{q}}}_{q = q_E}}
\
\ee
leading to
\be\label{eq:Gamma}
\Gamma \sim W \min\lrp{1, W\abs{\frac{d q}{d j_{q}}}_{q = q_E}} \sim W N^{-b}, \quad b\geq 0 \ .
\ee
In the latter equality of~\eqref{eq:DOS_sum} we substitute the expression for the spectrum $j_q$, Eq.~\eqref{eq:j_q}, to Eq.~\eqref{eq:ImG_Lorenz} and used the solution $q_E$ of the equation $j_{q_E} = E$, while the latter equality of Eq.~\eqref{eq:Gamma} should be considered as the definition of the scaling parameter $b$.

In addition, the above substitution determines
the crossover length scale $R_*$, separating the specific short-range behavior from the universal power-law decay.
For $a<1$ this crossover length scale takes the form
\be
R_* = \lrb{\frac{\Gamma^2+E^2}{t^2}}^{\frac{1}{2(1-a)}}
\ee
and at large distances $R\gg R_*$, the effective power-law decay switches to
\be\label{eq:j_eff_a<1}
j_{R}^{\rm eff} \sim  \frac{\Gamma}{t g(E)R^{2-a}} \ ,
\ee
In the opposite case of $a>1$ at $R\gg R_*$ only the prefactor may change with respect to the bare hopping, but not the power-law decay
\be\label{eq:j_eff_a>1}
j_{R}^{\rm eff} \sim  \frac{\Gamma t}{g(E)(E^2+\Gamma^2)R^{a}} \ .
\ee

Next, in order to understand the short-range structure of eigenstates, one should compare
the maximal $J^*$ and typical $j_q^{typ}$ hopping, Eq.~\eqref{eq:j_q}, as well as their level spacing with the broadening $\Gamma$.
Here and further we will focus on the scaling with $N$ in order to distinguish ergodic, fractal, and localized states.
Those who is not interested in the overall analysis of the cases, but would like to focus on the bulk-spectrum delocalization with the Poisson statistics, can directly go to the item~\ref{item:3}:
\begin{enumerate}
  \item In the case of large broadening $\Gamma \gg J^*>|j_q|$, corresponding to $\gamma+2b<\min\lrp{0, 2a-2}$,
  one obtains in the bulk the Anderson localization with $g(E) \sim 1/W$, $\Gamma\sim W$, $b=0$, and
the non-rescaled result at $\gamma<\min\lrp{0, 2a-2}\leq 0$
\be
j_{R}^{\rm eff} \sim  \frac{\Gamma W \theta(W/2-|E|)}{(E^2+\Gamma^2)}\frac{t}{R^{a}} \sim \frac{t}{R^{a}} \
\ee
similarly to the usual BM-model at $a>1$.

  \item In the case when the broadening is in between the maximal and typical $j_q$, $j_q^{typ}\ll \Gamma\ll J^*$, corresponding to $2a-2<\gamma<-2b\leq 0$ (i.e., $a<1$) the Green's function has two contributions
\be\label{eq:G_0_2sums}
{\bar G}_{mm}(E)=\frac{1}{N}\sum_q \frac{1}{ E - \Sigma-j_q} \simeq f \frac{1}{E-\Sigma} + \frac{1}{N}\sum_{j_q>\Gamma} \frac{1}{ E - j_q}
 \ .
\ee
The first one is reminiscent to the box DOS, with
\be
f = 1 - \frac{q_\Gamma}{\pi}, \quad j_{q_\Gamma} = \Gamma \ ,
\ee
being the fraction of $q$ with $j_q<\Gamma$,
while the second term gives the high-energy tail of the DOS.
Here $1/N\gg q_* \gg 1$ is given by
\be
q_* = \lrp{\frac{t}{\Gamma}}^{1/(1-a)} \sim N^{-\frac{|\gamma|-2b}{2(1-a)}} \ ,
\ee
thus, $f \to 1$ in the thermodynamic limit.
Consider separately the bulk and the edge spectral states.

In the bulk of the spectrum, $|E|<\Gamma$, the second term in~\eqref{eq:G_0_2sums} is not resonant $j_q>\Gamma>E$ and is given by
\be
\frac{1}{N}\sum_{j_q>\Gamma} \frac{1}{j_q} \sim \frac{1}{N t}\sum_{q<q_*} q^{1-a} \sim \frac{1}{N \Gamma} \sim N^{b-1} \ .
\ee
Assuming that $b<1$ one will obtain the usual box term to be dominant giving again $\Gamma\sim W$, with $b=0$.
The effective hopping in this case
\be\label{eq:j_eff_MI}
j_{R}^{\rm eff} \sim  \min\lrb{\frac{t}{R^a},\frac{W^2}{t R^{2-a}}} \ .
\ee


At the edge of the spectrum $|E|\gg W$ the main contribution is given by the corresponding pole $E = j_q$ at $q=q_E$ and gives the result $g(E) \simeq \abs{{d q}/{d j_{q}}}_{q = q_E} \ll 1/W$ and $\Gamma \sim W^2 g(E)$.
This pole results in the hopping term
\be
j_{R}^{\rm eff} \simeq \frac{t}{R^a}\frac{W^2}{\lrp{E - t R^{1-a}}^2+\lrp{\frac{W^2 q_E^{2-a}}{t}}^2} \ ,
\ee
with the resonance at the extensive distance $R\sim 1/q_E\gg 1$.
The value at the resonance is very sensitive to the position of the energy $E$ between the adjacent levels $j_{1/R}$, with $R$ closest to $1/q_E$.
Thus the variable
\be
\delta E = E - t R^{1-a}
\ee
varies  in the range
\be
|\delta E|\lesssim \Delta E = |j_{1/R}-j_{1/(R+1)}| \sim t R^{-a} \sim t q_E^a\ .
\ee

As soon as $|E|\gg W$, $\Delta E\gg \Gamma$ and the average resonance hopping takes the form
\be\label{eq:j_R^res}
j_{\rm res} = \mean{j_R^{eff,res}} = \int_{-\Delta E/2}^{\Delta E/2} \frac{d\delta E}{\Delta E} j_{R}^{\rm eff} \simeq \int_{-\infty}^{\infty} \frac{W^2 d\delta E}{\delta E^2+\lrp{\frac{W^2 q_E^{2-a}}{t}}^2} \sim t q_E^{a-2}
\ee

In addition, as $\Delta E\gg \Gamma$ such hopping term is resonant for the {\it only site} from the initial site, which is located at the distance $R \sim 1/q_E$.
As a result, such a resonant consideration is equivalent to the effective $1$d Anderson model with the exponential localization
\be\label{eq:psi_exp_loc}
|\psi(R)|^2\sim \frac{e^{-|R-R_0|/\xi}}{\xi} \ ,
\ee
and the localization length given by the resonant hopping $j_{\rm res}$
\be\label{eq:loc_length}
\xi \sim \lrp{\frac{j_{\rm res}}{W}}^2 \sim \lrp{\frac{t q_E^{a-2}}{W}}^2\gg N^{\frac{|\gamma|}{1-a}} \ ,
\ee
as $|E|\gg W$ and, thus, $q_E\ll q_* = N^{-\frac{|\gamma|}{2(1-a)}}$.
Note that here the normalization condition of the wave function is taken into account by the prefactor $1/\xi$.

As soon as localization length $\xi$ scales up with the system size $N$, the corresponding eigenstates at the spectral edge are delocalized.
These states become ergodic, when the localization length becomes of the order of the system size, $\xi \sim N^{1}$, i.e at~\footnote{Note that for $\gamma = 0$ this coincides with the edge where the eigenstates become localized in the momentum space~\cite{Nosov2019correlation}.}
\be\label{eq:q_*}
q_E \lesssim \lrp{N^{1/2}/t}^{-1/(2-a)} = N^{-\frac{1+|\gamma|}{2(2-a)}} \ .
\ee

The states in the middle
$
N^{-\frac{1+|\gamma|}{2(2-a)}}\ll q_E \lesssim N^{-\frac{|\gamma|}{2(1-a)}}
$,
i.e., at $O(1)\lesssim E\ll N^{\frac{1-a-|\gamma|}{2(2-a)}}$
are delocalized, but non-ergodic. This works as well for the conventional Burin-Maksimov model, $\gamma = 0$, however, as mentioned in the introduction,
such non-ergodic delocalized states form a measure zero of all states.

\item \label{item:3}In the most interesting case when the broadening is much smaller than the typical hopping, but much larger than its level spacing, $\delta j_q^{typ}\ll \Gamma\ll j_q^{typ}$, corresponding to $0<\gamma+2b<2$, the fractal properties of the edge states come to the bulk of the spectrum.
Indeed, on one hand, in Eq.~\eqref{eq:G_0_2sums} the fraction $f\to 0$ goes to zero in the thermodynamic limit and, thus, the density of states in the spectral bulk is given by the pole contributions, $g(E) \simeq \abs{{d q}/{d j_{q}}}_{q = q_E} \ll 1/W$ and $\Gamma \sim W^2 g(E)$.

On the other hand, the broadening makes this spectrum continuous and the effective hopping reads as
\be\label{eq:j_R_result}
j_{R}^{\rm eff} \simeq \frac{t}{R^a}\frac{W^2}{\lrp{E - j_{1/R}}^2+\Gamma^2} \ ,
\ee
has not only the resonance hopping like~\eqref{eq:j_R^res} at $R\simeq 1/q_E$, but also the small power-law tail at $R\gg 1/q_E$ from $j_{1/R} \sim t R^{1-a}$~\eqref{eq:j_eff_a<1} for $a<1$
\be\label{eq:j_R_result_a<1}
j_{R}^{\rm eff} \sim  \frac{W^2}{t R^{2-a}} \ ,
\ee
and~\eqref{eq:j_eff_a>1} for $a>1$
\be\label{eq:j_R_result_a>1}
j_{R}^{\rm eff} \sim  \frac{W^2 t}{\lrp{E^2+\lrp{\frac{W^2 q_E^{2-a}}{t}}^2}R^{a}}\lesssim  \frac{W^2}{t R^{a}} \ .
\ee
Both these power-law terms are small compared to the diagonal disorder $W$ and, thus, perturbative and cannot lead to delocalization.

Eventually in the bulk of the spectrum $E\sim t$, $q_E\sim N^0 \ll 1$ and, thus, $\Gamma \sim W^2/t$, corresponding to $b = -\gamma/2$.
This limits the corresponding case to the range
\be
0<\gamma<1 \ .
\ee

However, the presence of resonant hops $j_{\rm res} \sim t$ at $R\simeq 1/q_E \sim O(1)$ should delocalize the wave function similarly to the edge spectral states in the previous item.
Indeed, the resonances form the short-range Anderson model with the random hopping terms on top of the weak rescaled power-law tails.
Thus, we should have the exponentially decaying wave function~\eqref{eq:psi_exp_loc} with the $N$-dependent localization length~\eqref{eq:loc_length}
\be\label{eq:loc_length_bulk}
\xi \sim \lrp{\frac{j_{\rm res}}{W}}^2 \sim \lrp{\frac{t q_E^{a-2}}{W}}^2\sim N^{\gamma} \ .
\ee
This localization length exceeds the system size for $\gamma>1$ and leads to the localization at $\gamma<0$ in agreement with the considered range of $\gamma$ and the properties of the bulk spectral states in the previous items.

%
%
%

At the top part of the spectrum similarly to the previous section there is the ergodic-nonergodic crossover in energy giving the same effective momentum~\eqref{eq:q_*}.
The bottom of the spectrum where $j_q \sim t(\pi -q)^2$ is also a bit special, but we skip its consideration here for brevity.

\item In the last case when $\Gamma\ll \delta j_q^{typ}$ most of the states are just localized in the momentum space.
\end{enumerate}

The above analytical results for the case~\ref{item:3}, $0<\gamma<1$, can be summarized as follows:
both for $a<1$ and $a>1$ the effective hopping term~\eqref{eq:j_R_result} after renormalization combines the resonances at $j_{1/R} \simeq E$
leading to the exponential decay~\eqref{eq:psi_exp_loc} of the bulk eigenstates with the localization length~\eqref{eq:xi_N^gamma} growing with the system size $N$.
At larger distances, $R\gtrsim \xi\sim N^{\gamma}$, the hopping~\eqref{eq:j_R_result} becomes power-law decaying and perturbative giving for $a<1$ Eq.~\eqref{eq:j_R_result_a<1} and for $a>1$ Eq.~\eqref{eq:j_R_result_a>1}
\be
j_{R}^{\rm eff} \sim  \frac{W^2}{t R^{a_{\rm eff}}} \sim \frac{W}{N^{\gamma/2} R^{a_{\rm eff}}} \ , \quad a_{\rm eff} = \max\lrp{a,2-a} \ ,
\ee
similarly to the case of the conventional BM model~\eqref{eq:psi_decay}.
This agrees very well with the numerical simulations of the typical wave-function decay, Eq.~\eqref{eq:psi_typ_decay}, shown in Figs.~\ref{fig:PL-decay} and~\ref{fig:exp-decay}.

The above exponential~\eqref{eq:psi_exp_loc} and the power-law~\eqref{eq:j_R_result} decay result in the piecewise linear $f(\alpha)$, Eq.~\eqref{eq:f(alpha)_def}.
The exponential decay with the prefactor $1/\xi$ corresponds to the plateau at $r\ll \xi$, where $|\psi_E(r)|\sim 1/\xi$
\be
f(\alpha) = \frac{\ln \xi}{\ln N} = \gamma \ ,
\ee
and to a sharp cutoff at $r>\xi$, corresponding to $\alpha<\gamma$.
The power-law decay leads to the linear increase
\be
f(\alpha_R) = \frac{\ln R}{\ln N} = \frac{\alpha_R-\gamma}{2 a_{\rm eff}}, \quad \alpha_R = -\frac{\ln j_R^{\rm eff}}{\ln N} \ .
\ee
This result is in fair agreement with Fig.~\ref{fig:f(a)}.

The absence of the level repulsion, shown in Fig.~\ref{fig:r-stat} is confirmed by the analytical consideration of the resonances.
Indeed, as effectively the model including only resonance couplings is equivalent to the $1$d Anderson model with the hopping on the distance $\sim 1/q_E\gg 1$,
it effectively separates the spectrum into $N q_E\ll N$ blocks of the size $1/q_E$ and therefore leads to the Poisson level statistics.


\section{Conclusion and outlook}
In this paper, we develop the framework in the random-matrix setting, which mimics the properties of the eigenstates in the many-body localized phase, namely,
the non-ergodic delocalization of the wave functions in the Hilbert space combined with the Poisson level statistics.
For this, we formulate the ingredients needed for the realization of the above wave-function structure in random-matrix ensembles and apply them to a concrete model, relevant for the experimental observation in trapped ions, ultra cold and Rydberg atoms.

Namely, we consider a so-called Burin-Maksimov model with the long-range correlated hopping term which does not lead to the delocalization of the bulk spectral states, due to the renormalization of the power-law tails of the hopping, similar somehow to the structure of the local integrals of motion emerging in the many-body localized phase of disordered interacting models.
We generalize this model to the case of the non-ergodic delocalized eigenstates in the bulk of the spectrum, by rescaling the hopping term with the matrix size $N$ (mimicking the Hilbert space dimension).
The developed analytical approach allows us to investigate the eigenstate statistics and spatial profile and to find the structure of resonances in the effective hopping.
It is shown that the eigenstates both for short- and long-range models,
first, decay exponentially with the localization length $\xi$, scaling up with $N$, and then go to the power-law decay. It is this localization length scale-up, $\xi \sim N^{\gamma}$, which makes the wave function non-ergodically delocalized for $0<\gamma<1$ and forms effectively the extensive number of blocks of repulsing eigenvalues.
The latter leads to the Poisson statistics.
The numerical simulations explicitly confirm the above analytical investigations.

These results provide the way to mimic the wave-function structure in the Hilbert space of the many-body localized phase of matter
with rather simple random matrix ensembles
and open many possibilities to investigate further properties of the eigenstates using such probes as
the subdiffusive wave-packet spreading in the Hilbert space (see, e.g.,~\cite{Biroli2017delocalized,Bera2018return,DeTomasi2019subdiffusion,Biroli2020anomalous,Nosov2022statistics}),
the logarithmic entanglement growth in the static and driven setting~\cite{Bardarson_Sent_log(t)},
the relation of the entanglement to non-ergodicity~\cite{Fradkin_Sq_Dq,DeTomasi_Sq_Dq_2020} or even out-of-time-ordered correlators~\cite{Fan_OTOC_MBL}.
Although the presence of the extensive number of ergodic eigenstates at the spectral edge move our model closer to the many-body mobility edge, present just below the MBL phase, the zero fraction of these states makes our model more suitable for the description of the very vicinity of the MBL transition and finite-temperature MBL for smaller disorder amplitudes.

The more deep understanding of global spectral properties would be also very helpful in order to understand how to distinguish
the Poisson statistics of the localized and non-ergodic delocalized wave functions.

\begin{acknowledgements}
We are grateful to V.~E.~Kravtsov, A.~G.~Kutlin, and A.~L.~Burin for insightful discussions.
The work of I.~M.~K. is supported by the Russian Science Foundation under the grant 21-12-00409.
\end{acknowledgements}
\bibliography{Lib}

\begin{thebibliography}{73}
\providecommand{\natexlab}[1]{#1}
\providecommand{\url}[1]{\texttt{#1}}
\expandafter\ifx\csname urlstyle\endcsname\relax
  \providecommand{\doi}[1]{doi: #1}\else
  \providecommand{\doi}{doi: \begingroup \urlstyle{rm}\Url}\fi

\bibitem[Abanin et~al.(2019)Abanin, Altman, Bloch, and Serbyn]{Abanin_RMP}
D.~A. Abanin, E.~Altman, I.~Bloch, and M.~Serbyn.
\newblock Colloquium: Many-body localization, thermalization, and entanglement.
\newblock \emph{Rev. Mod. Phys.}, 91:\penalty0 021001, May 2019.
\newblock \doi{10.1103/RevModPhys.91.021001}.
\newblock URL \url{https://doi.org/10.1103/RevModPhys.91.021001}.

\bibitem[Alet and Laflorencie(2018)]{Alet_CRP}
F.~Alet and N.~Laflorencie.
\newblock Many-body localization: An introduction and selected topics.
\newblock \emph{Comptes Rendus Physique}, 19\penalty0 (6):\penalty0 498 -- 525,
  2018.
\newblock ISSN 1631-0705.
\newblock \doi{10.1016/j.crhy.2018.03.003}.
\newblock URL \url{https://doi.org/10.1016/j.crhy.2018.03.003}.
\newblock Quantum simulation / Simulation quantique.

\bibitem[Atas et~al.(2013{\natexlab{a}})Atas, Bogomolny, Giraud, and
  Roux]{Atas2013distribution}
Y.~Y. Atas, E.~Bogomolny, O.~Giraud, and G.~Roux.
\newblock Distribution of the ratio of consecutive level spacings in random
  matrix ensembles.
\newblock \emph{Phys. Rev. Lett.}, 110:\penalty0 084101, 2013{\natexlab{a}}.
\newblock \doi{10.1103/PhysRevLett.110.084101}.
\newblock URL \url{https://doi.org/10.1103/PhysRevLett.110.084101}.

\bibitem[Atas et~al.(2013{\natexlab{b}})Atas, Bogomolny, Giraud, Vivo, and
  Vivo]{Atas2013joint}
Y.~Y. Atas, E.~Bogomolny, O.~Giraud, P.~Vivo, and E.~Vivo.
\newblock Joint probability densities of level spacing ratios in random
  matrices.
\newblock \emph{Journal of Physics A: Mathematical and Theoretical},
  46\penalty0 (35):\penalty0 355204, aug 2013{\natexlab{b}}.
\newblock \doi{10.1088/1751-8113/46/35/355204}.
\newblock URL \url{https://doi.org/10.1088%2F1751-8113%2F46%2F35%2F355204}.

\bibitem[Bardarson et~al.(2012)Bardarson, Pollmann, and
  Moore]{Bardarson_Sent_log(t)}
J.~H. Bardarson, F.~Pollmann, and J.~E. Moore.
\newblock Unbounded growth of entanglement in models of many-body localization.
\newblock \emph{Phys. Rev. Lett.}, 109:\penalty0 017202, Jul 2012.
\newblock \doi{10.1103/PhysRevLett.109.017202}.
\newblock URL \url{https://doi.org/10.1103/PhysRevLett.109.017202}.

\bibitem[Basko et~al.(2006)Basko, Aleiner, and Altshuler]{Basko06}
D.~M. Basko, I.~L. Aleiner, and B.~L. Altshuler.
\newblock Metal-insulator transition in a weakly interacting many-electron
  system with localized single-particle states.
\newblock \emph{Annals of Physics}, 321\penalty0 (5):\penalty0 1126 -- 1205,
  2006.
\newblock ISSN 0003-4916.
\newblock \doi{10.1016/j.aop.2005.11.014}.
\newblock URL \url{https://doi.org/10.1016/j.aop.2005.11.014}.

\bibitem[Bera et~al.(2018)Bera, {De Tomasi}, Khaymovich, and
  Scardicchio]{Bera2018return}
S.~Bera, G.~{De Tomasi}, I.~M. Khaymovich, and A.~Scardicchio.
\newblock Return probability for the {Anderson} model on the random regular
  graph.
\newblock \emph{Phys. Rev. B}, 98:\penalty0 134205, 2018.
\newblock \doi{10.1103/PhysRevB.98.134205}.
\newblock URL \url{https://doi.org/10.1103/PhysRevB.98.134205}.

\bibitem[Berkovits(2020)]{Berkovits2020super}
R.~Berkovits.
\newblock Super-{Poissonian} behavior of the {Rosenzweig}-{Porter} model in the
  nonergodic extended regime.
\newblock \emph{Phys. Rev. B}, 102:\penalty0 165140, Oct 2020.
\newblock \doi{10.1103/PhysRevB.102.165140}.
\newblock URL \url{https://doi.org/10.1103/PhysRevB.102.165140}.

\bibitem[Berkovits(2021)]{Berkovits2021probing}
R.~Berkovits.
\newblock Probing the metallic energy spectrum beyond the thouless energy scale
  using singular value decomposition.
\newblock \emph{Phys. Rev. B}, 104:\penalty0 054207, Aug 2021.
\newblock \doi{10.1103/PhysRevB.104.054207}.
\newblock URL \url{https://doi.org/10.1103/PhysRevB.104.054207}.

\bibitem[Biroli and Tarzia(2017)]{Biroli2017delocalized}
G.~Biroli and M.~Tarzia.
\newblock Delocalized glassy dynamics and many-body localization.
\newblock \emph{Phys. Rev. B}, 96:\penalty0 201114(R), Nov 2017.
\newblock \doi{10.1103/PhysRevB.96.201114}.
\newblock URL \url{https://doi.org/10.1103/PhysRevB.96.201114}.

\bibitem[Biroli and Tarzia(2020)]{Biroli2020anomalous}
G.~Biroli and M.~Tarzia.
\newblock Anomalous dynamics on the ergodic side of the many-body localization
  transition and the glassy phase of directed polymers in random media.
\newblock \emph{Phys. Rev. B}, 102:\penalty0 064211, Aug 2020.
\newblock \doi{10.1103/PhysRevB.102.064211}.
\newblock URL \url{https://doi.org/10.1103/PhysRevB.102.064211}.

\bibitem[Biroli and Tarzia(2021)]{BirTar_Levy-RP}
G.~Biroli and M.~Tarzia.
\newblock {L}\'evy-{R}osenzweig-{P}orter random matrix ensemble.
\newblock \emph{Phys. Rev. B}, 103:\penalty0 104205, Mar 2021.
\newblock \doi{10.1103/PhysRevB.103.104205}.
\newblock URL \url{https://doi.org/10.1103/PhysRevB.103.104205}.

\bibitem[Burin(2017)]{Burin_AdP}
A.~Burin.
\newblock Localization and chaos in a quantum spin glass model in random
  longitudinal fields: Mapping to the localization problem in a {Bethe} lattice
  with a correlated disorder.
\newblock \emph{Annalen der Physik}, 529\penalty0 (7):\penalty0 1600292, 2017.
\newblock \doi{10.1002/andp.201600292}.
\newblock URL \url{https://doi.org/10.1002/andp.201600292}.

\bibitem[Burin(2015)]{BurinPRB2015-1}
A.~L. Burin.
\newblock Many-body delocalization in a strongly disordered system with
  long-range interactions: Finite-size scaling.
\newblock \emph{Phys. Rev. B}, 91:\penalty0 094202, 2015.
\newblock \doi{10.1103/PhysRevB.91.094202}.
\newblock URL \url{https://doi.org/10.1103/PhysRevB.91.094202}.

\bibitem[Burin and Maksimov(1989)]{Burin1989}
A.~L. Burin and L.~A. Maksimov.
\newblock Localization and delocalization of particles in disordered lattice
  with tunneling amplitude with $r^{-3}$ decay.
\newblock \emph{JETP Lett.}, 50:\penalty0 338, 1989.
\newblock URL \url{http://jetpletters.ru/ps/1129/article_17116.shtml}.

\bibitem[Celardo et~al.(2016)Celardo, Kaiser, and Borgonovi]{Borgonovi_2016}
G.~L. Celardo, R.~Kaiser, and F.~Borgonovi.
\newblock Shielding and localization in the presence of long-range hopping.
\newblock \emph{Phys. Rev. B}, 94:\penalty0 144206, 2016.
\newblock \doi{10.1103/PhysRevB.94.144206}.
\newblock URL \url{https://doi.org/10.1103/PhysRevB.94.144206}.

\bibitem[Chen et~al.(2015)Chen, Yu, Cho, Clark, and Fradkin]{Fradkin_Sq_Dq}
X.~Chen, X.~Yu, G.~Y. Cho, B.~K. Clark, and E.~Fradkin.
\newblock Many-body localization transition in {Rokhsar-Kivelson}-type wave
  functions.
\newblock \emph{Phys. Rev. B}, 92:\penalty0 214204, Dec 2015.
\newblock \doi{10.1103/PhysRevB.92.214204}.
\newblock URL \url{https://doi.org/10.1103/PhysRevB.92.214204}.

\bibitem[Colmenarez et~al.(2022)Colmenarez, Luitz, Khaymovich, and
  De~Tomasi]{colmenarez2021spectral}
L.~Colmenarez, D.~J. Luitz, I.~M. Khaymovich, and G.~De~Tomasi.
\newblock Subdiffusive thouless time scaling in the anderson model on random
  regular graphs.
\newblock \emph{Phys. Rev. B}, 105:\penalty0 174207, May 2022.
\newblock \doi{10.1103/PhysRevB.105.174207}.
\newblock URL \url{https://doi.org/10.1103/PhysRevB.105.174207}.

\bibitem[D'Alessio et~al.(2016)D'Alessio, Kafri, Polkovnikov, and
  Rigol]{DAlessio2016ETH}
L.~D'Alessio, Y.~Kafri, A.~Polkovnikov, and M.~Rigol.
\newblock From quantum chaos and eigenstate thermalization to statistical
  mechanics and thermodynamics.
\newblock \emph{Advances in Physics}, 65\penalty0 (3):\penalty0 239--362, 2016.
\newblock \doi{10.1080/00018732.2016.1198134}.
\newblock URL \url{http://dx.doi.org/10.1080/00018732.2016.1198134}.

\bibitem[de~Moura et~al.(2005)de~Moura, Malyshev, Lyra, Malyshev, and
  Dominguez-Adame]{Malyshev2005}
F.~A. B.~F. de~Moura, A.~V. Malyshev, M.~L. Lyra, V.~A. Malyshev, and
  F.~Dominguez-Adame.
\newblock Localization properties of a one-dimensional tight-binding model with
  nonrandom long-range intersite interactions.
\newblock \emph{Phys. Rev. B}, 71:\penalty0 174203, 2005.
\newblock \doi{10.1103/PhysRevB.71.174203}.
\newblock URL \url{https://doi.org/10.1103/PhysRevB.71.174203}.

\bibitem[De~Tomasi and Khaymovich(2020)]{DeTomasi_Sq_Dq_2020}
G.~De~Tomasi and I.~M. Khaymovich.
\newblock Multifractality meets entanglement: Relation for nonergodic extended
  states.
\newblock \emph{Phys. Rev. Lett.}, 124:\penalty0 200602, May 2020.
\newblock \doi{10.1103/PhysRevLett.124.200602}.
\newblock URL \url{https://doi.org/10.1103/PhysRevLett.124.200602}.

\bibitem[{De Tomasi} et~al.(2020){De Tomasi}, Bera, Scardicchio, and
  Khaymovich]{DeTomasi2019subdiffusion}
G.~{De Tomasi}, S.~Bera, A.~Scardicchio, and I.~M. Khaymovich.
\newblock Subdiffusion in the {Anderson} model on the random regular graph.
\newblock \emph{Phys. Rev. B}, 101:\penalty0 100201(R), Mar 2020.
\newblock \doi{10.1103/PhysRevB.101.100201}.
\newblock URL \url{https://doi.org/10.1103/PhysRevB.101.100201}.

\bibitem[De~Tomasi et~al.(2021)De~Tomasi, Khaymovich, Pollmann, and
  Warzel]{QIsing_2021}
G.~De~Tomasi, I.~M. Khaymovich, F.~Pollmann, and S.~Warzel.
\newblock Rare thermal bubbles at the many-body localization transition from
  the {Fock} space point of view.
\newblock \emph{Phys. Rev. B}, 104:\penalty0 024202, Jul 2021.
\newblock \doi{10.1103/PhysRevB.104.024202}.
\newblock URL \url{https://doi.org/10.1103/PhysRevB.104.024202}.

\bibitem[De~Tomasi and Khaymovich(2022)]{DeTomasi2022nonHermitian}
Giuseppe De~Tomasi and Ivan~M. Khaymovich.
\newblock Non-hermitian rosenzweig-porter random-matrix ensemble: Obstruction
  to the fractal phase.
\newblock \emph{Phys. Rev. B}, 106:\penalty0 094204, Sep 2022.
\newblock \doi{10.1103/PhysRevB.106.094204}.
\newblock URL \url{https://link.aps.org/doi/10.1103/PhysRevB.106.094204}.

\bibitem[Deng et~al.(2018)Deng, Kravtsov, Shlyapnikov, and
  Santos]{Deng2018duality}
X.~Deng, V.~E. Kravtsov, G.~V. Shlyapnikov, and L.~Santos.
\newblock Duality in power-law localization in disordered one-dimensional
  systems.
\newblock \emph{Phys. Rev. Lett.}, 120\penalty0 (11):\penalty0 110602, 2018.
\newblock \doi{10.1103/PhysRevLett.120.110602}.
\newblock URL \url{https://doi.org/10.1103/PhysRevLett.120.110602}.

\bibitem[Deng et~al.(2022)Deng, Burin, and Khaymovich]{deng2020anisotropy}
Xiaolong Deng, Alexander~L. Burin, and Ivan~M. Khaymovich.
\newblock {Anisotropy-mediated reentrant localization}.
\newblock \emph{SciPost Phys.}, 13:\penalty0 116, 2022.
\newblock \doi{10.21468/SciPostPhys.13.5.116}.
\newblock URL \url{https://scipost.org/10.21468/SciPostPhys.13.5.116}.

\bibitem[Deutsch(1991)]{Deutsch1991}
J.~M. Deutsch.
\newblock Quantum statistical mechanics in a closed system.
\newblock \emph{Phys. Rev. A}, 43:\penalty0 2046--2049, Feb 1991.
\newblock \doi{10.1103/PhysRevA.43.2046}.
\newblock URL \url{https://doi.org/10.1103/PhysRevA.43.2046}.

\bibitem[Evers and Mirlin(2008)]{Evers2008anderson}
F.~Evers and A.~D. Mirlin.
\newblock {Anderson} transitions.
\newblock \emph{Rev. Mod. Phys}, 80:\penalty0 1355, 2008.
\newblock \doi{10.1103/RevModPhys.80.1355}.
\newblock URL \url{https://doi.org/10.1103/RevModPhys.80.1355}.

\bibitem[Fan et~al.(2017)Fan, Zhang, Shen, and Zhai]{Fan_OTOC_MBL}
R.~Fan, P.~Zhang, H.~Shen, and H.~Zhai.
\newblock Out-of-time-order correlation for many-body localization.
\newblock \emph{Science Bulletin}, 62\penalty0 (10):\penalty0 707--711, 2017.
\newblock ISSN 2095-9273.
\newblock \doi{10.1016/j.scib.2017.04.011}.
\newblock URL \url{https://doi.org/10.1016/j.scib.2017.04.011}.

\bibitem[Gornyi et~al.(2005)Gornyi, Mirlin, and
  Polyakov]{gornyi2005interacting}
I.~V. Gornyi, A.~D. Mirlin, and D.~G. Polyakov.
\newblock Interacting electrons in disordered wires: {Anderson} localization
  and low-$t$ transport.
\newblock \emph{Phys. Rev. Lett.}, 95:\penalty0 206603, Nov 2005.
\newblock \doi{10.1103/PhysRevLett.95.206603}.
\newblock URL \url{https://doi.org/10.1103/PhysRevLett.95.206603}.

\bibitem[Haque et~al.(2022)Haque, McClarty, and Khaymovich]{Sent2020_Haque}
M.~Haque, P.~A. McClarty, and I.~M. Khaymovich.
\newblock Entanglement of midspectrum eigenstates of chaotic many-body systems:
  Reasons for deviation from random ensembles.
\newblock \emph{Phys. Rev. E}, 105:\penalty0 014109, Jan 2022.
\newblock \doi{10.1103/PhysRevE.105.014109}.
\newblock URL \url{https://doi.org/10.1103/PhysRevE.105.014109}.

\bibitem[Huang et~al.(2016)Huang, Zhang, and Chen]{Huang_OTOC_MBL}
Y.~Huang, Y.-L. Zhang, and X.~Chen.
\newblock Out-of-time-ordered correlators in many-body localized systems.
\newblock \emph{Annalen der Physik}, 529\penalty0 (7):\penalty0 1600318, 2016.
\newblock \doi{10.1002/andp.201600318}.
\newblock URL \url{https://doi.org/10.1002/andp.201600318}.

\bibitem[Huse et~al.(2014)Huse, Nandkishore, and
  Oganesyan]{huse2014phenomenology}
D.~A. Huse, R.~Nandkishore, and V.~Oganesyan.
\newblock Phenomenology of fully many-body-localized systems.
\newblock \emph{Phys. Rev. B}, 90:\penalty0 174202, Nov 2014.
\newblock \doi{10.1103/PhysRevB.90.174202}.
\newblock URL \url{https://doi.org/10.1103/PhysRevB.90.174202}.

\bibitem[Khaymovich and Kravtsov(2021)]{LN-RP_dyn}
I.~M. Khaymovich and V.~E. Kravtsov.
\newblock Dynamical phases in a ``multifractal'' {Rosenzweig}-{Porter} model.
\newblock \emph{SciPost Phys.}, 11:\penalty0 45, 2021.
\newblock \doi{10.21468/SciPostPhys.11.2.045}.
\newblock URL \url{https://doi.org/10.21468/SciPostPhys.11.2.045}.

\bibitem[Khaymovich et~al.(2020)Khaymovich, Kravtsov, Altshuler, and
  Ioffe]{LN-RP_WE}
I.~M. Khaymovich, V.~E. Kravtsov, B.~L. Altshuler, and L.~B. Ioffe.
\newblock Fragile ergodic phases in logarithmically-normal
  {Rosenzweig}-{{Porter}} model.
\newblock \emph{Phys. Rev. Research}, 2:\penalty0 043346, 2020.
\newblock \doi{10.1103/PhysRevResearch.2.043346}.
\newblock URL \url{https://doi.org/10.1103/PhysRevResearch.2.043346}.

\bibitem[Kravtsov et~al.(2015)Kravtsov, Khaymovich, Cuevas, and
  Amini]{Kravtsov_NJP2015}
V.~E. Kravtsov, I.~M. Khaymovich, E.~Cuevas, and M.~Amini.
\newblock A random matrix model with localization and ergodic transitions.
\newblock \emph{New J. Phys.}, 17:\penalty0 122002, 2015.
\newblock \doi{10.1088/1367-2630/17/12/122002}.
\newblock URL \url{https://doi.org/10.1088%2F1367-2630%2F17%2F12%2F122002}.

\bibitem[Kravtsov et~al.(2020)Kravtsov, Khaymovich, Altshuler, and
  Ioffe]{LN-RP_RRG}
V.~E. Kravtsov, I.~M. Khaymovich, B.~L. Altshuler, and L.~B. Ioffe.
\newblock Localization transition on the random regular graph as an unstable
  tricritical point in a log-normal rosenzweig-porter random matrix ensemble,
  2020.
\newblock URL \url{https://arxiv.org/abs/2002.02979}.

\bibitem[Kutlin and Khaymovich(2020)]{Kutlin2020_PLE-RG}
A.~G. Kutlin and I.~M. Khaymovich.
\newblock Renormalization to localization without a small parameter.
\newblock \emph{SciPost Phys.}, 8:\penalty0 49, 2020.
\newblock \doi{10.21468/SciPostPhys.8.4.049}.
\newblock URL \url{https://doi.org/10.21468/SciPostPhys.8.4.049}.

\bibitem[Kutlin and Khaymovich(2021)]{Kutlin2021emergent}
A.~G. Kutlin and I.~M. Khaymovich.
\newblock Emergent fractal phase in energy stratified random models.
\newblock \emph{SciPost Phys.}, 11:\penalty0 101, 2021.
\newblock \doi{10.21468/SciPostPhys.11.6.101}.
\newblock URL \url{https://doi.org/10.21468/SciPostPhys.11.6.101}.

\bibitem[Kutlin and Khaymovich(2023)]{kutlin2023anatomy}
A.~G. Kutlin and I.~M. Khaymovich.
\newblock Anatomy of the eigenstates distribution: a quest for a genuine
  multifractality, 2023.
\newblock URL \url{https://arxiv.org/abs/2309.06468}.

\bibitem[Levitov(1989)]{Levitov1989}
L.~S. Levitov.
\newblock Absence of localization of vibrational modes due to dipole-dipole
  interaction.
\newblock \emph{Europhys. Lett.}, 9:\penalty0 83, 1989.
\newblock \doi{10.1209/0295-5075/9/1/015}.
\newblock URL \url{https://doi.org/10.1209%2F0295-5075%2F9%2F1%2F015}.

\bibitem[Levitov(1990)]{Levitov1990}
L.~S. Levitov.
\newblock Delocalization of vibrational modes caused by electric dipole
  interaction.
\newblock \emph{Phys. Rev. Lett.}, 64:\penalty0 547, 1990.
\newblock \doi{10.1103/PhysRevLett.64.547}.
\newblock URL \url{https://doi.org/10.1103/PhysRevLett.64.547}.

\bibitem[Luitz et~al.(2015)Luitz, Laflorencie, and Alet]{Luitz15}
D.~J. Luitz, N.~Laflorencie, and F.~Alet.
\newblock Many-body localization edge in the random-field {H}eisenberg chain.
\newblock \emph{Phys. Rev. B}, 91:\penalty0 081103, Feb 2015.
\newblock \doi{10.1103/PhysRevB.91.081103}.
\newblock URL \url{https://doi.org/10.1103/PhysRevB.91.081103}.

\bibitem[Luitz et~al.(2020)Luitz, Khaymovich, and {Bar
  Lev}]{luitz2019multifractality}
D.~J. Luitz, I.~M. Khaymovich, and Y.~{Bar Lev}.
\newblock {Multifractality and its role in anomalous transport in the
  disordered {XXZ} spin-chain}.
\newblock \emph{SciPost Phys. Core}, 2:\penalty0 6, 2020.
\newblock \doi{10.21468/SciPostPhysCore.2.2.006}.
\newblock URL \url{https://doi.org/10.21468/SciPostPhysCore.2.2.006}.

\bibitem[Mac\'e et~al.(2019)Mac\'e, Alet, and
  Laflorencie]{Mace_Laflorencie2019_XXZ}
N.~Mac\'e, F.~Alet, and N.~Laflorencie.
\newblock Multifractal scalings across the many-body localization transition.
\newblock \emph{Phys. Rev. Lett.}, 123:\penalty0 180601, Oct 2019.
\newblock \doi{10.1103/PhysRevLett.123.180601}.
\newblock URL \url{https://doi.org/10.1103/PhysRevLett.123.180601}.

\bibitem[Mehta(2004)]{Mehta2004random}
M.~L. Mehta.
\newblock \emph{Random matrices}.
\newblock Elsevier, 2004.
\newblock \doi{10.1016/C2009-0-22297-5}.
\newblock URL \url{https://doi.org/10.1016/C2009-0-22297-5}.

\bibitem[Mirlin et~al.(1996)Mirlin, Fyodorov, Dittes, Quezada, and
  Seligman]{MirFyod1996}
A.~D. Mirlin, Y.~V. Fyodorov, F.-M. Dittes, J.~Quezada, and T.~H. Seligman.
\newblock Transition from localized to extended eigenstates in the ensemble of
  power-law random banded matrices.
\newblock \emph{Phys. Rev. E}, 54:\penalty0 3221, 1996.
\newblock \doi{10.1103/PhysRevE.54.3221}.
\newblock URL \url{https://doi.org/10.1103/PhysRevE.54.3221}.

\bibitem[Morningstar et~al.(2022)Morningstar, Colmenarez, Khemani, Luitz, and
  Huse]{Huse21}
A.~Morningstar, L.~Colmenarez, V.~Khemani, D.~J. Luitz, and D.~A. Huse.
\newblock Avalanches and many-body resonances in many-body localized systems.
\newblock \emph{Phys. Rev. B}, 105:\penalty0 174205, May 2022.
\newblock \doi{10.1103/PhysRevB.105.174205}.
\newblock URL \url{https://doi.org/10.1103/PhysRevB.105.174205}.

\bibitem[Motamarri et~al.(2022)Motamarri, Gorsky, and
  Khaymovich]{Motamarri2021RDM}
Vedant~R. Motamarri, Alexander~S. Gorsky, and Ivan~M. Khaymovich.
\newblock {Localization and fractality in disordered Russian Doll model}.
\newblock \emph{SciPost Phys.}, 13:\penalty0 117, 2022.
\newblock \doi{10.21468/SciPostPhys.13.5.117}.
\newblock URL \url{https://scipost.org/10.21468/SciPostPhys.13.5.117}.

\bibitem[Nosov and Khaymovich(2019)]{Nosov2019mixtures}
P.~A. Nosov and I.~M. Khaymovich.
\newblock Robustness of delocalization to the inclusion of soft constraints in
  long-range random models.
\newblock \emph{Phys. Rev. B}, 99:\penalty0 224208, Jun 2019.
\newblock \doi{10.1103/PhysRevB.99.224208}.
\newblock URL \url{https://doi.org/10.1103/PhysRevB.99.224208}.

\bibitem[Nosov et~al.(2019)Nosov, Khaymovich, and
  Kravtsov]{Nosov2019correlation}
P.~A. Nosov, I.~M. Khaymovich, and V.~E. Kravtsov.
\newblock Correlation-induced localization.
\newblock \emph{Physical Review B}, 99\penalty0 (10):\penalty0 104203, 2019.
\newblock \doi{10.1103/PhysRevB.99.104203}.
\newblock URL \url{https://doi.org/10.1103/PhysRevB.99.104203}.

\bibitem[Nosov et~al.(2022)Nosov, Khaymovich, Kudlis, and
  Kravtsov]{Nosov2022statistics}
P.~A. Nosov, I.~M. Khaymovich, A.~Kudlis, and V.~E. Kravtsov.
\newblock {Statistics of Green's functions on a disordered Cayley tree and the
  validity of forward scattering approximation}.
\newblock \emph{SciPost Phys.}, 12:\penalty0 48, 2022.
\newblock \doi{10.21468/SciPostPhys.12.2.048}.
\newblock URL \url{https://doi.org/10.21468/SciPostPhys.12.2.048}.

\bibitem[Oganesyan and Huse(2007)]{oganesyan2007localization}
V.~Oganesyan and D.~A. Huse.
\newblock Localization of interacting fermions at high temperature.
\newblock \emph{Phys. Rev. B}, 75:\penalty0 155111, Apr 2007.
\newblock \doi{10.1103/PhysRevB.75.155111}.
\newblock URL \url{https://doi.org/10.1103/PhysRevB.75.155111}.

\bibitem[Pal and Huse(2010)]{Pal2010}
A.~Pal and D.~A. Huse.
\newblock Many-body localization phase transition.
\newblock \emph{Phys. Rev. B}, 82:\penalty0 174411, Nov 2010.
\newblock \doi{10.1103/PhysRevB.82.174411}.
\newblock URL \url{https://doi.org/10.1103/PhysRevB.82.174411}.

\bibitem[Polkovnikov et~al.(2011)Polkovnikov, Sengupta, Silva, and
  Vengalattore]{Polkovnikov_2011}
A.~Polkovnikov, K.~Sengupta, A.~Silva, and M.~Vengalattore.
\newblock Colloquium: Nonequilibrium dynamics of closed interacting quantum
  systems.
\newblock \emph{Rev. Mod. Phys.}, 83:\penalty0 863--883, Aug 2011.
\newblock \doi{10.1103/RevModPhys.83.863}.
\newblock URL \url{https://doi.org/10.1103/RevModPhys.83.863}.

\bibitem[Ray et~al.(2018)Ray, Ghosh, and Sinha]{Ray_Floquet_MF}
S.~Ray, A.~Ghosh, and S.~Sinha.
\newblock Drive-induced delocalization in the {Aubry}-{A}ndr\'e model.
\newblock \emph{Phys. Rev. E}, 97:\penalty0 010101, Jan 2018.
\newblock \doi{10.1103/PhysRevE.97.010101}.
\newblock URL \url{https://doi.org/10.1103/PhysRevE.97.010101}.

\bibitem[Rigol et~al.(2008)Rigol, Dunjko, and
  Olshanii]{rigol2008thermalization}
M.~Rigol, V.~Dunjko, and M.~Olshanii.
\newblock Thermalization and its mechanism for generic isolated quantum
  systems.
\newblock \emph{Nature}, 452\penalty0 (7189):\penalty0 854, apr 2008.
\newblock \doi{10.1038/nature06838}.
\newblock URL \url{https://doi.org/10.1038/nature06838}.

\bibitem[Riser and Kanzieper(2021)]{Riser2021power}
R.~Riser and E.~Kanzieper.
\newblock Power spectrum and form factor in random diagonal matrices and
  integrable billiards.
\newblock \emph{Annals of Physics}, 425:\penalty0 168393, 2021.
\newblock ISSN 0003-4916.
\newblock \doi{10.1016/j.aop.2020.168393}.
\newblock URL \url{https://doi.org/10.1016/j.aop.2020.168393}.

\bibitem[Riser et~al.(2017)Riser, Osipov, and Kanzieper]{Riser2017power}
R.~Riser, V.~Al. Osipov, and E.~Kanzieper.
\newblock Power spectrum of long eigenlevel sequences in quantum chaotic
  systems.
\newblock \emph{Phys. Rev. Lett.}, 118:\penalty0 204101, May 2017.
\newblock \doi{10.1103/PhysRevLett.118.204101}.
\newblock URL \url{https://doi.org/10.1103/PhysRevLett.118.204101}.

\bibitem[Riser et~al.(2020)Riser, Osipov, and
  Kanzieper]{Riser2020nonperturbative}
R.~Riser, V.~Al. Osipov, and E.~Kanzieper.
\newblock Nonperturbative theory of power spectrum in complex systems.
\newblock \emph{Annals of Physics}, 413:\penalty0 168065, 2020.
\newblock ISSN 0003-4916.
\newblock \doi{10.1016/j.aop.2019.168065}.
\newblock URL \url{https://doi.org/10.1016/j.aop.2019.168065}.

\bibitem[{Rosenzweig} and {Porter}(1960)]{RP}
N.~{Rosenzweig} and C.~E. {Porter}.
\newblock "repulsion of energy levels" in complex atomic spectra.
\newblock \emph{Phys. Rev. B}, 120:\penalty0 1698, 1960.
\newblock \doi{10.1103/PhysRev.120.1698}.
\newblock URL \url{https://doi.org/10.1103/PhysRev.120.1698}.

\bibitem[Roy et~al.(2018)Roy, Khaymovich, Das, and Moessner]{Floquet_MF}
S.~Roy, I.~M. Khaymovich, A.~Das, and R.~Moessner.
\newblock {Multifractality without fine-tuning in a {Floquet} quasiperiodic
  chain}.
\newblock \emph{SciPost Phys.}, 4:\penalty0 25, 2018.
\newblock \doi{10.21468/SciPostPhys.4.5.025}.
\newblock URL \url{https://doi.org/10.21468/SciPostPhys.4.5.025}.

\bibitem[Sarkar et~al.(2021)Sarkar, Ghosh, Sen, and
  Sengupta]{Sarkar_Floquet_MF}
M.~Sarkar, R.~Ghosh, A.~Sen, and K.~Sengupta.
\newblock Mobility edge and multifractality in a periodically driven
  {Aubry-A}ndr\'e model.
\newblock \emph{Phys. Rev. B}, 103:\penalty0 184309, May 2021.
\newblock \doi{10.1103/PhysRevB.103.184309}.
\newblock URL \url{https://doi.org/10.1103/PhysRevB.103.184309}.

\bibitem[Sarkar et~al.(2022)Sarkar, Ghosh, Sen, and
  Sengupta]{Sarkar2021signatures}
M.~Sarkar, R.~Ghosh, A.~Sen, and K.~Sengupta.
\newblock Signatures of multifractality in a periodically driven interacting
  aubry-andr\'e model.
\newblock \emph{Phys. Rev. B}, 105:\penalty0 024301, Jan 2022.
\newblock \doi{10.1103/PhysRevB.105.024301}.
\newblock URL \url{https://doi.org/10.1103/PhysRevB.105.024301}.

\bibitem[Serbyn et~al.(2013)Serbyn, Papi\ifmmode~\acute{c}\else \'{c}\fi{}, and
  Abanin]{serbyn2013local}
M.~Serbyn, Z.~Papi\ifmmode~\acute{c}\else \'{c}\fi{}, and D.~A. Abanin.
\newblock Local conservation laws and the structure of the many-body localized
  states.
\newblock \emph{Phys. Rev. Lett.}, 111:\penalty0 127201, Sep 2013.
\newblock \doi{10.1103/PhysRevLett.111.127201}.
\newblock URL \url{https://doi.org/10.1103/PhysRevLett.111.127201}.

\bibitem[Srednicki(1994)]{Srednicki1994}
M.~Srednicki.
\newblock Chaos and quantum thermalization.
\newblock \emph{Phys. Rev. E}, 50:\penalty0 888--901, Aug 1994.
\newblock \doi{10.1103/PhysRevE.50.888}.
\newblock URL \url{https://doi.org/10.1103/PhysRevE.50.888}.

\bibitem[Srednicki(1996)]{Srednicki1996}
M.~Srednicki.
\newblock Thermal fluctuations in quantized chaotic systems.
\newblock \emph{J. Phys. A: Mathematical and General}, 29\penalty0
  (4):\penalty0 L75, 1996.
\newblock \doi{10.1088/0305-4470/29/4/003}.
\newblock URL \url{https://doi.org/10.1088/0305-4470/29/4/003}.

\bibitem[Tarzia(2020)]{Tarzia_2020}
M.~Tarzia.
\newblock Many-body localization transition in {Hilbert space}.
\newblock \emph{Phys. Rev. B}, 102:\penalty0 014208, Jul 2020.
\newblock \doi{10.1103/PhysRevB.102.014208}.
\newblock URL \url{https://doi.org/10.1103/PhysRevB.102.014208}.

\bibitem[Tekur et~al.(2018)Tekur, Bhosale, and Santhanam]{Tekur2018higher}
S.~H. Tekur, U.~T. Bhosale, and M.~S. Santhanam.
\newblock Higher-order spacing ratios in random matrix theory and complex
  quantum systems.
\newblock \emph{Phys. Rev. B}, 98:\penalty0 104305, Sep 2018.
\newblock \doi{10.1103/PhysRevB.98.104305}.
\newblock URL \url{https://doi.org/10.1103/PhysRevB.98.104305}.

\bibitem[Tikhonov and Mirlin(2018)]{Tikhonov2018MBL_long-range}
K.~S. Tikhonov and A.~D. Mirlin.
\newblock Many-body localization transition with power-law interactions:
  Statistics of eigenstates.
\newblock \emph{Phys. Rev. B}, 97:\penalty0 214205, Jun 2018.
\newblock \doi{10.1103/PhysRevB.97.214205}.
\newblock URL \url{https://doi.org/10.1103/PhysRevB.97.214205}.

\bibitem[Torres-Vargas et~al.(2017)Torres-Vargas, Fossion, Tapia-Ignacio, and
  L\'opez-Vieyra]{TorresVargas2017determination}
G.~Torres-Vargas, R.~Fossion, C.~Tapia-Ignacio, and J.~C. L\'opez-Vieyra.
\newblock Determination of scale invariance in random-matrix spectral
  fluctuations without unfolding.
\newblock \emph{Phys. Rev. E}, 96:\penalty0 012110, Jul 2017.
\newblock \doi{10.1103/PhysRevE.96.012110}.
\newblock URL \url{https://doi.org/10.1103/PhysRevE.96.012110}.

\bibitem[Torres-Vargas et~al.(2018)Torres-Vargas, M\'endez-Berm\'udez,
  L\'opez-Vieyra, and Fossion]{TorresVargas2018crossover}
G.~Torres-Vargas, J.~A. M\'endez-Berm\'udez, J.~C. L\'opez-Vieyra, and
  R.~Fossion.
\newblock Crossover in nonstandard random-matrix spectral fluctuations without
  unfolding.
\newblock \emph{Phys. Rev. E}, 98:\penalty0 022110, Aug 2018.
\newblock \doi{10.1103/PhysRevE.98.022110}.
\newblock URL \url{https://doi.org/10.1103/PhysRevE.98.022110}.

\bibitem[Yonezawa and Morigaki(1973)]{Yonezawa1973coherent_potential}
F.~Yonezawa and K.~Morigaki.
\newblock Coherent potential approximation. {Basic} concepts and applications.
\newblock \emph{Progress of Theoretical Physics Supplement}, 53:\penalty0
  1--76, 01 1973.
\newblock ISSN 0375-9687.
\newblock \doi{10.1143/PTPS.53.1}.
\newblock URL \url{https://doi.org/10.1143/PTPS.53.1}.

\end{thebibliography}

\appendix

\section{Coherent potential approximation}\label{AppSec:Coherent_potential}
In this section we derive a coherent potential approximation for
the Green's function $\hat G(E)$ solving
the problem
\be\label{eq:G(E)}
(E-\hat H)\hat G(E) =  \hat 1 \ ,
\ee
with the Hamiltonian
\be\label{eq:ham_App}
\hat H = \hat \ep+\hat j \
\ee
represented by
\be
\hat \ep = \sum_n \ep_n \lv n \ra \la n \rv
\ee
the diagonal disorder in the coordinate basis $\{\lv n \ra\}$, with
\be
\la \ep_n \ra = 0, \quad \la \ep_n^2\ra = W^2 \ ,
\ee
 and
\be
\hat j = \sum_{m,n} j_{mn}\ket{m}\bra{n} = \sum_{q} j_{q}\ket{q} \bra{q}
\ee
hopping term diagonalized in a certain basis $\{\ket{q}\}$.

In order to solve the problem we first assume that after averaging over $\hat \ep$ the self-energy part $\Sigma$ is a scalar
\be\label{eq:G_Sigma}
(E-\Sigma-\hat j)\hat{\bar G}(E) =  \hat 1 \Leftrightarrow  \hat{\bar G} =
\sum_{q} \frac{\ket{q} \bra{q}}{E-\Sigma -j_{q}} \ .
\ee

Then this self-energy can be found self-consistently assuming the disorder is averaged everywhere except one site $k$:
\be
\lb E-\Sigma-\hat j - (\ep_k - \Sigma)\lv k \ra \la k \rv\rb\hat G(E,k) =  \hat 1 \ .
\ee
Here $\hat G(E,k) = \la \hat G(E)\ra_{\ep_{n\ne k}}$.
Taking into account left part of \eqref{eq:G_Sigma} one can get
\be\label{eq:G_ep_k_eq}
\lb \hat 1 - \hat{\bar G} (\ep_k - \Sigma)\lv k \ra \la k \rv\rb\hat G(E,k) =  \hat{\bar G}(E) \
\ee
or after multiplying by $\la k \rv$
\be\label{eq:<k|G_ep_k_eq}
\lb 1 - (\ep_k - \Sigma) {\bar G}_{kk}(E) \rb  \la k \rv\hat G(E,k) =  \la k \rv\hat{\bar G}(E) \Leftrightarrow
\la k \rv\hat G(E,k) =  \frac{\la k \rv\hat{\bar G}(E)}{1 - (\ep_k - \Sigma) {\bar G}_{kk}(E) } \ ,
\ee
where ${\bar G}_{kk}(E)=\la k \rv\hat{\bar G}(E)\lv k \ra$. Note that ${\bar G}_{kk}(E) \equiv {\bar G}_{0}(E)$ does not depend on $k$ if the basis $\{\lv q \ra\}$ momentum basis $|\la n| q \ra|^2 = 1/N$, where $N$ is the size of the system.
Averaging the latter equation over a certain distribution $P(\ep)$ of $\ep=\ep_k$ and dividing by a common factor $\la k \rv\hat{\bar G}(E)$
we get the self-consistency equation
\be\label{eq:self-consist}
\int \frac{P(\ep)d\ep}{1 - (\ep - \Sigma) {\bar G}_{0}(E) } = 1
\ee
Decomposing $1$ into $\int P(\ep)d\ep$ we get
\be\label{eq:self-consist_Sigma}
\int P(\ep)d\ep\frac{\ep - \Sigma}{1 - (\ep - \Sigma) {\bar G}_{0}(E) } = 0 \Leftrightarrow
\Sigma = \int \frac{\ep P(\ep)d\ep}{1 - (\ep - \Sigma) {\bar G}_{0}(E) }
\ee
In the left equation we divided by ${\bar G}_{0}(E)$, while in the right one we take into account~\eqref{eq:self-consist}.

Eventually, substituting the right equation of~\eqref{eq:self-consist_Sigma} to the right one of~\eqref{eq:G_Sigma}, we obtain
\be\label{eq:G_Sigma_res}
\hat{\bar G} = \sum_{q} \frac{\lv q \ra \la q \rv}{E-\int \frac{\ep P(\ep)d\ep}{1 - (\ep - \Sigma) {\bar G}_{0}(E) } -j_{q}} \ .
\ee

\begin{itemize}
  \item For the Lorenzian distribution $P(\ep) = (W/\pi)[\ep^2+W^2]^{-1}$ we get from~\eqref{eq:self-consist}
\be\label{eq:Sigma_Lorenz}
1 = \int \frac{W}{\pi(\ep^2+W^2)}\frac{d\ep}{1 - (\ep - \Sigma) {\bar G}_{0}(E) } = \frac{1}{1 + (i W + \Sigma) {\bar G}_{0}(E) }
\ee
leading to $\Sigma = -i W$ for any ${\bar G}_{0}(E)$.

  \item On the other hand, the box distribution leads to
\begin{multline}\label{eq:Sigma_Box}
W{\bar G}_{0}(E) = \int_{-W/2}^{W/2} \frac{d\ep}{{\bar G}_{0}^{-1}(E)+ \Sigma - \ep } = \ln\frac{1 + (\Sigma+W/2) {\bar G}_{0}(E)}{1 + (\Sigma-W/2) {\bar G}_{0}(E) } \lra\\ \Sigma = -{\bar G}_{0}^{-1}(E) +\frac{W}{2\tanh \lb W {\bar G}_{0}(E)/2\rb} \ ,
\end{multline}
giving the result for the self-energy
\be\label{eq:Sigma_Box_limits}
\Sigma = \left\{
           \begin{array}{ll}
             \frac{W^2}{12}{\bar G}_{0}(E), & W|{\bar G}_{0}(E)|\ll 1 \\
             \frac{W}{2} -{\bar G}_{0}^{-1}(E), & W|{\bar G}_{0}(E)|\gg 1
           \end{array}
         \right.
\ee
and the level broadening
\be\label{eq:Gamma_limits}
\Gamma = \Im\Sigma = \left\{
           \begin{array}{ll}
             \pi \frac{W^2}{12}g(E), & W|{\bar G}_{0}(E)|\ll 1 \\
             \frac{\pi g(E)}{|{\bar G}_{0}(E)|^2}, & W|{\bar G}_{0}(E)|\gg 1
           \end{array}
         \right. = \pi g(E) \min\lrp{\frac{W^2}{12}, \frac{1}{|{\bar G}_{0}(E)|^2}} \ .
\ee
The first result is compatible with the Fermi Golden rule as in this case $\Im \Sigma =  \pi \la \ep_n^2\ra g(E)$, $\la \ep_n^2\ra = W^2/12$.
\end{itemize}

The global DOS is given by
\be\label{App_eq:DOS_sum}
g(E) = \frac{\Im {\bar G}_{0}(E)}{\pi} = \frac{1}{\pi N}\sum_q \frac{\Gamma}{(E - j_q - \Re \Sigma)^2 + \Gamma^2} \ .
\ee


In order to verify our results, let's consider the limiting cases
\begin{enumerate}
  \item First, let's consider $|j_q|\ll \max(E,\Gamma)$.
Then from the second equality in~\eqref{eq:G_Sigma} and the definition of ${\bar G}_{0}(E)=\la k \rv\hat{\bar G}(E)\lv k \ra$ one immediately obtains
\be
{\bar G}_{0}^{-1}(E) = E - \Sigma
\ee
and substituting it to the integral~\eqref{eq:Sigma_Box} gets
\be
{\bar G}_{0}(E)=\frac{1}{W}\ln\frac{E+W/2}{E -W/2} \ .
\ee
As the argument of the logarithm is real for $|E|>W/2$, the DOS $g(E)$ as the imaginary part of ${\bar G}_{0}(E)$ is zero.
Otherwise, for $|E|<W$ the negative sign of the argument provides the additional imaginary part of the logarithm given by $\pi$ and
results in the obvious result
\be\label{eq:DOS_1/W}
g(E) = \frac{\Im {\bar G}_{0}(E)}{\pi} = \frac{\theta(W/2-|E|)}{W} \ .
\ee
Note that in Eq.~\eqref{eq:DOS_sum} there is always the trivial solution $g(E)$ which might affect the numerical calculations.
In this case
\be\label{eq:Gamma_box}
\Gamma = \frac{\pi W \theta\lrp{W/2 - |E|}}{\lrp{\ln\lrv{\frac{E+W/2}{E -W/2}}}^2+\pi^2} \sim W N^0  \ .
\ee
In the opposite case of large typical $|j_q^{typ}|\gg \Gamma$ the DOS is given mostly by the poles at $E = j_q$.
  \item In the continuous case of $|j_q^{typ}|\gg \Gamma\gg \delta j_q^{typ}$
\begin{multline}\label{eq:G_0_dq/dj}
{\bar G}_{0}(E)=\frac{1}{N}\sum_q \frac{1}{ E - \Sigma-j_q} \simeq \int \frac{d q}{2\pi}\frac{1}{ E - j_q} = V.P.\int\frac{d q}{2\pi}\frac{1}{ E - j_q} + i\pi \subs{\frac{d q}{d j_q}}_{j_q = E}
\sim\\ \sim g(E)\lrp{\ln\lrv{\frac{J^*-E}{E}}+i\pi} \ ,
\end{multline}
with
\be\label{eq:DOS_dq/dj}
g(E) =  \subs{\frac{d q}{d j_{q}}}_{q = q_E} \ll \frac1W\ ,
\ee
and $q_E$ being the solution of the equation $j_{q_E} = E$.
Then the broadening is given by
\be\label{eq:Gamma_dq_dj}
\Gamma \simeq \frac{\pi W^2}{12} g(E) \sim W^2 \subs{\frac{d q}{d j_{q}}}_{q = q_E}
\ee
as $|{\bar G}_{0}(E)|\ll 1/W$.

\item In the last case of $|j_q^{typ}|\gg \delta j_q \gg \Gamma$ the density of states is given by the delta-peaks at the positions of the poles $E = j_q$.

\end{enumerate}

%
%


\end{document}